\documentstyle{elsart}
\newcommand{\der}[2]{\frac{\partial #1}{\partial #2}}
\newcommand{\imu}{\sqrt{-1}}
\expandafter\ifx\csname amssym.def\endcsname\relax \else\endinput\fi
\expandafter\edef\csname amssym.def\endcsname{%
       \catcode`\noexpand\@=\the\catcode`\@\space}
\catcode`\@=11
\def\undefine#1{\let#1\undefined}
\def\newsymbol#1#2#3#4#5{\let\next@\relax
 \ifnum#2=\@ne\let\next@\msafam@\else
 \ifnum#2=\tw@\let\next@\msbfam@\fi\fi
 \mathchardef#1="#3\next@#4#5}
\def\mathhexbox@#1#2#3{\relax
 \ifmmode\mathpalette{}{\m@th\mathchar"#1#2#3}%
 \else\leavevmode\hbox{$\m@th\mathchar"#1#2#3$}\fi}
\def\hexnumber@#1{\ifcase#1 0\or 1\or 2\or 3\or 4\or 5\or 6\or 7\or 8\or
 9\or A\or B\or C\or D\or E\or F\fi}

\font\tenmsa=msam10
\font\sevenmsa=msam7
\font\fivemsa=msam5
\newfam\msafam
\textfont\msafam=\tenmsa
\scriptfont\msafam=\sevenmsa
\scriptscriptfont\msafam=\fivemsa
\edef\msafam@{\hexnumber@\msafam}
\mathchardef\dabar@"0\msafam@39
\def\dashrightarrow{\mathrel{\dabar@\dabar@\mathchar"0\msafam@4B}}
\def\dashleftarrow{\mathrel{\mathchar"0\msafam@4C\dabar@\dabar@}}

\def\ulcorner{\delimiter"4\msafam@70\msafam@70 }
\def\urcorner{\delimiter"5\msafam@71\msafam@71 }
\def\llcorner{\delimiter"4\msafam@78\msafam@78 }
\def\lrcorner{\delimiter"5\msafam@79\msafam@79 }
\def\yen{{\mathhexbox@\msafam@55 }}
\def\checkmark{{\mathhexbox@\msafam@58 }}
\def\circledR{{\mathhexbox@\msafam@72 }}
\def\maltese{{\mathhexbox@\msafam@7A }}

\font\tenmsb=msbm10
\font\sevenmsb=msbm7
\font\fivemsb=msbm5
\newfam\msbfam
\textfont\msbfam=\tenmsb
\scriptfont\msbfam=\sevenmsb
\scriptscriptfont\msbfam=\fivemsb
\edef\msbfam@{\hexnumber@\msbfam}
\def\Bbb#1{{\fam\msbfam\relax#1}}
\def\widehat#1{\setbox\z@\hbox{$\m@th#1$}%
 \ifdim\wd\z@>\tw@ em\mathaccent"0\msbfam@5B{#1}%
 \else\mathaccent"0362{#1}\fi}
\def\widetilde#1{\setbox\z@\hbox{$\m@th#1$}%
 \ifdim\wd\z@>\tw@ em\mathaccent"0\msbfam@5D{#1}%
 \else\mathaccent"0365{#1}\fi}
\font\teneufm=eufm10
\font\seveneufm=eufm7
\font\fiveeufm=eufm5
\newfam\eufmfam
\textfont\eufmfam=\teneufm
\scriptfont\eufmfam=\seveneufm
\scriptscriptfont\eufmfam=\fiveeufm

\csname amssym.def\endcsname

\expandafter\ifx\csname pre amssym.tex at\endcsname\relax \else \endinput\fi
\expandafter\chardef\csname pre amssym.tex at\endcsname=\the\catcode`\@
\catcode`\@=11
\newsymbol\boxdot 1200
\newsymbol\boxplus 1201
\newsymbol\boxtimes 1202
\newsymbol\square 1003
\newsymbol\blacksquare 1004
\newsymbol\centerdot 1205
\newsymbol\lozenge 1006
\newsymbol\blacklozenge 1007
\newsymbol\circlearrowright 1308
\newsymbol\circlearrowleft 1309
\undefine\rightleftharpoons
\newsymbol\rightleftharpoons 130A
\newsymbol\leftrightharpoons 130B
\newsymbol\boxminus 120C
\newsymbol\Vdash 130D
\newsymbol\Vvdash 130E
\newsymbol\vDash 130F
\newsymbol\twoheadrightarrow 1310
\newsymbol\twoheadleftarrow 1311
\newsymbol\leftleftarrows 1312
\newsymbol\rightrightarrows 1313
\newsymbol\upuparrows 1314
\newsymbol\downdownarrows 1315
\newsymbol\upharpoonright 1316
 
\newsymbol\downharpoonright 1317
\newsymbol\upharpoonleft 1318
\newsymbol\downharpoonleft 1319
\newsymbol\rightarrowtail 131A
\newsymbol\leftarrowtail 131B
\newsymbol\leftrightarrows 131C
\newsymbol\rightleftarrows 131D
\newsymbol\Lsh 131E
\newsymbol\Rsh 131F
\newsymbol\rightsquigarrow 1320
\newsymbol\leftrightsquigarrow 1321
\newsymbol\looparrowleft 1322
\newsymbol\looparrowright 1323
\newsymbol\circeq 1324
\newsymbol\succsim 1325
\newsymbol\gtrsim 1326
\newsymbol\gtrapprox 1327
\newsymbol\multimap 1328
\newsymbol\therefore 1329
\newsymbol\because 132A
\newsymbol\doteqdot 132B
 
\newsymbol\triangleq 132C
\newsymbol\precsim 132D
\newsymbol\lesssim 132E
\newsymbol\lessapprox 132F
\newsymbol\eqslantless 1330
\newsymbol\eqslantgtr 1331
\newsymbol\curlyeqprec 1332
\newsymbol\curlyeqsucc 1333
\newsymbol\preccurlyeq 1334
\newsymbol\leqq 1335
\newsymbol\leqslant 1336
\newsymbol\lessgtr 1337
\newsymbol\backprime 1038
\newsymbol\risingdotseq 133A
\newsymbol\fallingdotseq 133B
\newsymbol\succcurlyeq 133C
\newsymbol\geqq 133D
\newsymbol\geqslant 133E
\newsymbol\gtrless 133F
\newsymbol\sqsubset 1340
\newsymbol\sqsupset 1341
\newsymbol\vartriangleright 1342
\newsymbol\vartriangleleft 1343
\newsymbol\trianglerighteq 1344
\newsymbol\trianglelefteq 1345
\newsymbol\bigstar 1046
\newsymbol\between 1347
\newsymbol\blacktriangledown 1048
\newsymbol\blacktriangleright 1349
\newsymbol\blacktriangleleft 134A
\newsymbol\vartriangle 134D
\newsymbol\blacktriangle 104E
\newsymbol\triangledown 104F
\newsymbol\eqcirc 1350
\newsymbol\lesseqgtr 1351
\newsymbol\gtreqless 1352
\newsymbol\lesseqqgtr 1353
\newsymbol\gtreqqless 1354
\newsymbol\Rrightarrow 1356
\newsymbol\Lleftarrow 1357
\newsymbol\veebar 1259
\newsymbol\barwedge 125A
\newsymbol\doublebarwedge 125B
\undefine\angle
\newsymbol\angle 105C
\newsymbol\measuredangle 105D
\newsymbol\sphericalangle 105E
\newsymbol\varpropto 135F
\newsymbol\smallsmile 1360
\newsymbol\smallfrown 1361
\newsymbol\Subset 1362
\newsymbol\Supset 1363
\newsymbol\Cup 1264
 
\newsymbol\Cap 1265
 
\newsymbol\curlywedge 1266
\newsymbol\curlyvee 1267
\newsymbol\leftthreetimes 1268
\newsymbol\rightthreetimes 1269
\newsymbol\subseteqq 136A
\newsymbol\supseteqq 136B
\newsymbol\bumpeq 136C
\newsymbol\Bumpeq 136D
\newsymbol\lll 136E
 
\newsymbol\ggg 136F
 
\newsymbol\circledS 1073
\newsymbol\pitchfork 1374
\newsymbol\dotplus 1275
\newsymbol\backsim 1376
\newsymbol\backsimeq 1377
\newsymbol\complement 107B
\newsymbol\intercal 127C
\newsymbol\circledcirc 127D
\newsymbol\circledast 127E
\newsymbol\circleddash 127F
\newsymbol\lvertneqq 2300
\newsymbol\gvertneqq 2301
\newsymbol\nleq 2302
\newsymbol\ngeq 2303
\newsymbol\nless 2304
\newsymbol\ngtr 2305
\newsymbol\nprec 2306
\newsymbol\nsucc 2307
\newsymbol\lneqq 2308
\newsymbol\gneqq 2309
\newsymbol\nleqslant 230A
\newsymbol\ngeqslant 230B
\newsymbol\lneq 230C
\newsymbol\gneq 230D
\newsymbol\npreceq 230E
\newsymbol\nsucceq 230F
\newsymbol\precnsim 2310
\newsymbol\succnsim 2311
\newsymbol\lnsim 2312
\newsymbol\gnsim 2313
\newsymbol\nleqq 2314
\newsymbol\ngeqq 2315
\newsymbol\precneqq 2316
\newsymbol\succneqq 2317
\newsymbol\precnapprox 2318
\newsymbol\succnapprox 2319
\newsymbol\lnapprox 231A
\newsymbol\gnapprox 231B
\newsymbol\nsim 231C
\newsymbol\ncong 231D
\newsymbol\diagup 231E
\newsymbol\diagdown 231F
\newsymbol\varsubsetneq 2320
\newsymbol\varsupsetneq 2321
\newsymbol\nsubseteqq 2322
\newsymbol\nsupseteqq 2323
\newsymbol\subsetneqq 2324
\newsymbol\supsetneqq 2325
\newsymbol\varsubsetneqq 2326
\newsymbol\varsupsetneqq 2327
\newsymbol\subsetneq 2328
\newsymbol\supsetneq 2329
\newsymbol\nsubseteq 232A
\newsymbol\nsupseteq 232B
\newsymbol\nparallel 232C
\newsymbol\nmid 232D
\newsymbol\nshortmid 232E
\newsymbol\nshortparallel 232F
\newsymbol\nvdash 2330
\newsymbol\nVdash 2331
\newsymbol\nvDash 2332
\newsymbol\nVDash 2333
\newsymbol\ntrianglerighteq 2334
\newsymbol\ntrianglelefteq 2335
\newsymbol\ntriangleleft 2336
\newsymbol\ntriangleright 2337
\newsymbol\nleftarrow 2338
\newsymbol\nrightarrow 2339
\newsymbol\nLeftarrow 233A
\newsymbol\nRightarrow 233B
\newsymbol\nLeftrightarrow 233C
\newsymbol\nleftrightarrow 233D
\newsymbol\divideontimes 223E
\newsymbol\varnothing 203F
\newsymbol\nexists 2040
\newsymbol\Finv 2060
\newsymbol\Game 2061
\newsymbol\mho 2066
\newsymbol\eth 2067
\newsymbol\eqsim 2368
\newsymbol\beth 2069
\newsymbol\gimel 206A
\newsymbol\daleth 206B
\newsymbol\lessdot 236C
\newsymbol\gtrdot 236D
\newsymbol\ltimes 226E
\newsymbol\rtimes 226F
\newsymbol\shortmid 2370
\newsymbol\shortparallel 2371
\newsymbol\smallsetminus 2272
\newsymbol\thicksim 2373
\newsymbol\thickapprox 2374
\newsymbol\approxeq 2375
\newsymbol\succapprox 2376
\newsymbol\precapprox 2377
\newsymbol\curvearrowleft 2378
\newsymbol\curvearrowright 2379
\newsymbol\digamma 207A
\newsymbol\varkappa 207B
\newsymbol\Bbbk 207C
\newsymbol\hslash 207D
\undefine\hbar
\newsymbol\hbar 207E
\newsymbol\backepsilon 237F
\catcode`\@=\csname pre amssym.tex at\endcsname

\newtheorem{Lemma}{Lemma}[section]
\newtheorem{Theorem}[Lemma]{Theorem}
\newtheorem{Corollary}[Lemma]{Corollary}

\newtheorem{Example}{Example}
\newtheorem{Remark}[Lemma]{Remark}
\begin{document}
\begin{frontmatter}
\title{Period preserving nonisospectral flows and the moduli space
of periodic solutions of soliton equations.}
\author{P.G.Grinevich\thanksref{G1}}
\thanks[G1]{The main part of this work was worked out during
the visit to the Freie Universit\"at of Berlin.
This visit was supported by the Volkswagen-Stiftung through the project
``Collaboration with scientists from the CIS''. This author was
also supported by the Soros International Scientific Foundation grant MD 8000
and by the Russian Foundation for Fundamental Studies grant 93-011-16087.}
\address{Landau Institute for Theoretical Physics,
Kosygina 2, Moscow, 117940, Russia. \\
e-mail: pgg@landau.ac.ru}
\and
\author{M.U.Schmidt\thanksref{S1}}
\thanks[S1]{Supported by DFG, SFB 288
'Differentialgeometrie and Quantenphysik'.}
\address{Institut f\"ur Theoretische Physik Freie Universit\"at Berlin,
Arnimallee 14, D14195 - Berlin, Germany. \\
e-mail:mschmidt@omega.physik.fu-berlin.de}
\begin{abstract}
Flows on the moduli space of the algebraic Riemann surfaces, preserving the
periods of the corresponding solutions of the soliton equations are
studied. We show that these flows are gradient with respect to some
indefinite symmetric flat metric arising in the Hamiltonian theory of the
Whitham equations. The functions generating these flows are conserved
quantities for all the equations simultaneously. We show that for 1+1
systems these flows can be imbedded in a larger system of ordinary nonlinear
differential equations with a rational right-hand side. Finally these flows
are used to give a complete description of the moduli space of
algebraic Riemann surfaces
corresponding to periodic solutions of the nonlinear Schr\"odinger equation.
\end{abstract}
\end{frontmatter}

\setcounter{section}{-1}
\section{Introduction.}

In the periodic theory of soliton equations we often meet the following
problem. The direct scattering transform is well-defined for purely periodic
potentials. In the quasiperiodic case the spectrum may have a much more
complicated structure (for example it may be similar to the Cantor set
\cite{DS}) and in general it is impossible to associate a Riemann surface to a
quasiperiodic potential. When we consider the inverse problem, general
algebraic Riemann surfaces generate quasiperiodic solutions.

Periods of solutions corresponding to a given algebraic Riemann surface
can be expressed in terms of some Abelian integrals. Thus the Riemann
surfaces corresponding to periodic solutions with a given period form
a transcendental submanifold in the moduli space of Riemann surfaces
of given genus. (More precisely we have to consider the moduli space
of Riemann surfaces of given genus with a set of marked points and jets of
local parameters in these points.)

In the present article we use the following approach to study these
submanifolds. It has been shown by one of the authors \cite{Sch}
that to any period preserving deformation corresponds a meromorphic
differential on the spectral curve with rather strong restrictions on
the singularities. Each such differential generates a flow on the
moduli space of Riemann surfaces leaving invariant the isoperiodic
submanifold. Earlier such flows were introduced by Krichever \cite{Kr1}
for more general meromorphic differentials in connection with the
topological quantum field theory and Whitham equations. In our paper
the period preserving flows are studied  more detailed in connection
with the problem of characterization of the periodic solutions,
especially for the $1+1$ systems like KdV and NLS.
Flows analogous to those studied in the present paper were introduced also by
Ercolani, Forest, McLaughlin and Shina \cite {EFMS} in the theory of
nonintegrable periodic perturbations of the soliton equations.

These flows can be naturally extended to general finite-gap solutions. In
this case they preserve the group of frequences.

In the $1+1$ case these flows can be written as ordinary differential
equations for the branch points but the right-hand side of of these
equations contains Abelian integrals. We show that these flows can be
essentially simplified by extending the configuration space. If we add
to the branch points the zeroes of the quasimomentum differential as
additional variables we obtain a system of ordinary differential
equations with a rational right-hand side. These flows are much more
convenient both for analytic studies and numerical simulations.
Differential equations on the zeroes of the quasimomentum differential
were written earlier in \cite{Kr1}. But our observation is  that
by combining these two sets of parameters together we get differential
equations with a rational right-hand side which are much simplier.

In some sense this system of ODE's is completely integrable, i.e.
it can be linearized in terms of some Abelian integrals. But it does not
mean integrability in the Painlev\'e sense because the structure of the
branch points may be rather complicated.

It is possible to introduce local coordinates on the moduli space given as
the values of the quasimomentum function in the stationary points of this
function. These coordinates are usually multivalued. Such coordinates
were also used by Krichever \cite{Kr1}, \cite{Kr2}, \cite{Kr4}.
In some specific cases like the real KdV and the
defocusing nonlinear Schr\"odinger equation
they are single-valued global coordinates. For the real KdV case similar
coordinates were used by Marchenko \cite{Mar}. In this case the moduli space
was also investigated by McKean and Van Moerbeke \cite{MKM} with the help
of a different coordinate system.

We show that if we change these coordinates the variations of the
corresponding Riemann surface are also described by ODE's with a
rational right-hand side. On the basis of these equations a numerical
program for studying isoperiodic deformations has been written\footnote
{The idea of the program interface was suggested to the authors
 by A. Bobenko.}.

Isoperiodic flows are rather naturally connected with the Whitham equations
in the Flaschka-Forest-McLaughlin form \cite{FFM}. The Hamiltonian theory of
the Whitham equations was constructed in some specific cases by Forest and
McLaughlin \cite{FM} in terms of action-angle variables and for general
situation by Dubrovin, Novikov \cite{DN} in the local differential-geometrical
form and later
developed by scientists from the Novikov's group. Novikov pointed out to
the authors that these flows should be gradient in the same flat Riemann
metric which arose in the Hamiltonian theory of the Whitham equations.
Here we give the proof of this conjecture.
A flow is called gradient if its vector field is defined as the gradient
of a function with respect to some symmetric Riemann metric. Let us recall
that Hamiltonian flows are defined as gradients of Hamiltonian functions
in a skew--symmetric metric.

In our case it is important that this symmetric metric is not positive
definite and even degenerate in some points. If we consider a gradient flow
in a positive definite metric it never conserves the function generating this
flow. But in our case the differentials of the functions generating
isoperiodic flows span a maximal isotropic subspace of our metric. Thus all
these functions are integrals of motion and our gradient system is similar
to an integrable Hamiltonian system. In fact there also exists a Hamiltonian
formulation of isoperiodic flows but the corresponding skew-symmetric form
is nonlocal with respect to the branch points.

We use this technique in the third part of our paper to study the moduli
space of all algebraic Riemann surfaces of finite genus generating periodic
solutions. We fix attention to the nonlinear Schr\"odinger equation and
show that there exists one-to-one correspondence between such Riemann
surfaces and some graphs together with the flow parameters mentioned above.
The multivaluedness of these parameters corresponds to different topologies
of these graphs. Using these graphs it is convenient to describe the branch
points of these coordinates. To formulate the hypotheses about the structure
of the moduli space proved later we used some numerical simulations.

\section {Flows preserving the periods.}

We shall study some nonisospectral flows on the space of the finite-gap
solutions of the soliton equations preserving the periods.

Let us recall some basic constructions from the periodic soliton theory
\cite{ZMNP}, \cite{Mu}. For definiteness let us consider the periodic
Korteweg-de Vries equation.

The $L$ operator for the Lax representation for KdV is the one-dimensional
Scr\"odinger operator
\begin{displaymath}
L=-\partial^2_x +u(x).
\end{displaymath}
If $u(x)$ is periodic with a period $P$ (without loss of generality we
shall assume $P=1$ such that $u(x+1)=u(x)$), then $L$ commutes with the shift
operator $S_x$ $(S_xf)(x)=f(x+1)$. The common eigenfunction $\Psi(\lambda,x)$
is called Bloch-Floquet function
\begin{displaymath}
L\Psi=\lambda\Psi,\ \Psi(\lambda,x+1)=\mu\Psi(\lambda,x).
\end{displaymath}
For generic $\lambda$ we have two eigenfunctions of $S_x$ with the
eigenvalues $\mu_+$ and $\mu_-$. In fact $\mu_+(\lambda)$ and $\mu_-(\lambda)$
are two branches of one function holomorphic on a hyperelliptic Riemann
surface $Y$, which is called the spectral curve. Finite-gap potentials
\cite{N1}
correspond to algebraic Riemann surfaces (i.e. surfaces with finite number
of branch points $\lambda_0$, $\lambda_1$, \ldots, $\lambda_{2g}$, $\infty$).
The function $p(\lambda)=\frac1{\imu}\ln(\mu)$ is called quasimomentum. It is
multivalued. The differential of the quasimomentum
$ dp $ is uniquely defined by the following properties:

\noindent 1) It is meromorphic on $Y$ with only one pole of second
order in the point $\infty$ $dp\sim\frac{d\lambda}{2\sqrt{\lambda}}$.

\noindent 2) All the periods of $dp$ are multiples of $2\pi$
\begin{displaymath}
\oint_{c_n}dp=2\pi k_n,\ k_n\in \Bbb Z
\end{displaymath}
over any closed cycle $c_n$.

In the inverse scattering problem we can consider the quasiperiodic
potentials as well as the periodic ones. We can start from a hyperelliptic
Riemann surface with arbitrary finite branch points $\lambda_0$, $\lambda_1$,
\ldots, $\lambda_{2g}$, ($\infty$ is a branch point too). The potentials
corresponding to this Riemann surface are parameterized by divisors on this
surface of degree $-g$ (or equivalently by collections of $g$ points
$\gamma_1$, \ldots, $\gamma_g$ in $Y$). In general these solutions are
quasiperiodic. The basic frequences are equal to the periods of the
differential $dp$, uniquely defined by the property 1) and

\noindent $2'$) All the periods of $dp$ are real:
\begin{displaymath}
\oint_{c_n}dp=2\pi k_n,\ k_n\in \Bbb R
\end{displaymath}
over any close cycle $c_n$.

The potential is periodic with period $P_*$ if and only if for all $n$
\begin{displaymath}
P_* k_n \in \Bbb Z.
\end{displaymath}

Thus we have the following characterization of the Riemann surfaces
corresponding to periodic potentials (the periodicity properties do not
depend on the choice of the divisor):

\noindent There exists a differential $dp$ with the properties 1) and 2).

We shall study deformations of Riemann surfaces preserving the periods of
$dp$. These deformations change the Riemann surfaces. Therefore we call them
nonisospectral, but they preserve all the $x$-frequences of the solutions.
In particular they leave invariant the space of potentials periodic in $x$
with a fixed period. Thus we call them isoperiodic.

For the real KdV condition $2'$) is equivalent to the to the following:

\noindent $2''$) All the $a$-periods of $dp$ are equal to zero
\begin{displaymath} \oint_{a_n} dp=0,\end{displaymath}
where $a_1$, \ldots, $a_g$
are the basic $a$-cycles, generated by real ovals. We can extend this
definition to the complex case and study deformations preserving
the $b$-periods of differential $dp$ with normalization $2''$) where
$a_1$, \ldots, $a_g$ are $a$-cycles from some fixed basis. In contrast
to $2'$) in the complex case this definition  depends on the choice of
the basis of cycles.

Let us consider the general situation.  We assume that the following
data are given:
\begin{description}
\item[(i)] a nonsingular compact Riemann surface \begin{math} Y\end{math},
\item[(ii)] a finite number of points \begin{math}
y_{1},\ldots ,y_{n}\end{math} on this Riemann surface and
\item[(iii)] local parameters \begin{math}
w_1=1/z_{1}\ldots ,w_n=1/z_{n} \end{math} near these
points.
\end{description}
Without loss of generality it is convenient to assume that
\begin{math} w_i\end{math}
is equal to zero at the point \begin{math} y_{i}\end{math}.

Now let \begin{math} H\end{math} be the group of all non-vanishing
holomorphic functions \begin{math} f\end{math}
defined on some set of the form \begin{math}
U\setminus \{ \infty \}\end{math}, where \begin{math}
U\subset {\Bbb {CP}}^{1}\end{math} is some
neighbourhood of \begin{math} \infty \end{math}
(which may depend on \begin{math} f\end{math}).
The group multiplication is given by multiplication of functions. The Lie
algebra of this group contains the vector space \begin{math} {\bf h}\end{math}
of all meromorphic functions on \begin{math}
{\Bbb {CP}}^{1}\end{math}, which are
holomorphic on \begin{math} {\Bbb C}\subset {\Bbb {CP}}^{1}\end{math}. With
the help of these data \begin{math} (Y,y_{1},\ldots ,y_{n},z_{1},\ldots
,z_{n})\end{math}, we define a group homomorphism
\begin{displaymath} L:H^{n}\rightarrow
\mbox{Picard group of \begin{math} Y\end{math}:}
\end{displaymath}
For this purpose we choose some covering \begin{displaymath} Y=Y\setminus
\{y_{1},\ldots ,y_{n}\} \cup U_{1} \cup \ldots \cup U_{n}
\end{displaymath}
{\tolerance=2000 with some small disjoint neighbourhoods \begin{math}
U_{1},\ldots ,U_{n}\end{math} of \begin{math}
y_{1},\ldots y_{n}\end{math}.
Then \begin{math} L(f_{1},\ldots ,f_{n})\end{math} is
defined with respect to this covering by
the cocycle, which is equal to \begin{math}
z_{i}\mapsto f(z_{i})\end{math} on \begin{math}
U_{i}\setminus \{ y_{i}\} \end{math} for all \begin{math} i=1,\ldots ,n
\end{math}. In particular  each element \begin{math} (f_{1},\ldots
,f_{n})\end{math} of \begin{math}
{\bf h}^{n}\end{math} defines a flow on the Picard
group of \begin{math} Y\end{math}, which
is given by multiplication with \begin{math}
L(\exp (tf_{1}),\ldots ,\exp (tf_{n}))\end{math}
for all \begin{math} t\in {\Bbb
C}\end{math}. These are the isospectral
flows of the dynamic systems, i.e. these flows do not change the
Riemann surface. In the soliton theory we have two fixed elements of
\begin{math} {\bf h}^{n}\end{math} generating
the spatial shifts and the flows generated by other elements are
associated with generalized time shifts. If one or both of the spatial
flows are trivial
we have a system with a reduced spatial dimension.
It is convenient to impose conditions on this flows, and
therefore to restrict the space of admissible data. In doing so the
dynamic system may become a Hamiltonian system. We will consider
three cases:

}
\begin{description}
\item[finite dimensional case:] The both fixed elements of
\begin{math} {\bf h}^{n} \end{math} induce
trivial flows on the Picard group of \begin{math} Y\end{math}.
In this case we have finite-dimensional ODE's and we do not need to discuss
periodicity properties. In this case the `isoperiodic deformations'
simply preserve the structure of these equations.
\item[simple periodic case:] One fixed
element of \begin{math} {\bf h}^{n}\end{math}
induces a trivial flow and another  fixed element of \begin{math}
{\bf h}^{n}\end{math}  induces a flow with period 1 on the Picard
group of \begin{math} Y\end{math}. In this case we have systems with
one spatial variable like the KdV or the nonlinear Schr\"odinger equation
and the isoperiodic deformations preserve the periodicity in this variable.
The meromorphic function generating the trivial flow maps our Riemann surface
\begin{math} Y\end{math} to the complex plane and it is natural to treat
\begin{math} Y\end{math} as a ramified covering of \begin{math} {\Bbb {CP}}^1
\end{math}. The positions of the branch points gives us the natural
coordinates on the moduli space.
\item[double periodic case:] The both fixed elements of
\begin{math} {\bf h}^{n}\end{math} induce flows with period
1 on the Picard group of \begin{math} Y\end{math}. We have systems with two
spatial variables like the Kadomtsev-Petviashvili equations and the
isoperiodic deformations preserve the periodicity in both spatial variables.
\end{description}
We assume that at all points \begin{math} y_{1},\ldots
,y_{n}\end{math} at least one of the two fixed elements
of \begin{math} {\bf h}^{n}\end{math} has a singularity. Otherwise we may
neglect the corresponding point and diminish the number
\begin{math} n\end{math} in our data. It is quite obvious that the flow
corresponding to some element \begin{math}
f\end{math} of \begin{math} {\bf h}^{n}\end{math}
is trivial, if and only if there exists some
meromorphic function on \begin{math}
Y\end{math}, which solves the Mittag Leffler distribution defined by the
cocycle corresponding to \begin{math} f\end{math}. This means that there
exists a function holomorphic on \begin{math}
Y\setminus \{ y_{1},\ldots y_{n}\} \end{math}
with prescribed singularities \begin{math} f_{i}(z_{i})+{\bf O}(1)\end{math}
in the points \begin{math} y_{i}\end{math},
\begin{math} i=1,\ldots n\end{math}. Furthermore, a
flow corresponding to some element \begin{math}
f\end{math} of \begin{math} {\bf h}^{n}\end{math}
has period 1 if and only if there exists some
non-vanishing holomorphic function on \begin{math}
Y\setminus \{ y_{1},\ldots y_{n}\}
\end{math}, such that the quotient of
this function divided by the function \begin{math}
z_{i}\mapsto \exp (f_{i}(z_{i}))\end{math}
extends to a holomorphic function near \begin{math} y_{i}\end{math} for all
\begin{math} i=1,\ldots n\end{math}.
Hence in the three cases mentioned above we have two
holomorphic functions \begin{math} \lambda \end{math}
and \begin{math} \mu \end{math} on \begin{math}
Y\setminus \{ y_{1},\ldots y_{n}\} \end{math}:
{\tolerance=2000 \begin{description}
\item[In the finite dimensional case]
both functions are meromorphic with
fixed Mittag Leffler distributions near \begin{math}
y_{1},\ldots ,y_{n}\end{math}.
\item[In the simple periodic case] \begin{math}
\lambda \end{math} is a meromorphic function
with fixed Mittag Leffler distribution
near \begin{math} y_{1},\ldots ,y_{n}\end{math}
and \begin{math} \mu \end{math} is a
non-vanishing holomorphic function on \begin{math} Y\setminus
\{y_{1},\ldots ,y_{n}\} \end{math}, such
that \begin{math} \ln (\mu )\end{math} has
a meromorphic branch with fixed Mittag
Leffler distribution near \begin{math}
y_{1},\ldots ,y_{n}\end{math}.
\item[In the double periodic case] both
functions \begin{math} \lambda \end{math} and \begin{math}
\mu \end{math} are non-vanishing holomorphic functions on
\begin{math} Y\setminus
\{y_{1},\ldots ,y_{n}\} \end{math}
, such that \begin{math} \ln
(\lambda )\end{math} and \begin{math} \ln
(\mu )\end{math} have
meromorphic branches with fixed Mittag
Leffler distributions near \begin{math}
y_{1},\ldots ,y_{n}\end{math}.
\end{description}
Hence in all cases these two functions
obey some relation of the form \begin{displaymath}
R(\lambda ,\mu )=0,\end{displaymath} where \begin{math}
R(\cdot ,\cdot ) \end{math} is some
holomorphic function on \begin{math}
{\Bbb C}\setminus \{ 0\} \times {\Bbb
C}\setminus \{ 0\} \end{math}. In
the finite dimensional case \begin{math}
R(\cdot ,\cdot )\end{math} may be
chosen to be a polynomial with respect
to \begin{math} \lambda \end{math}  and \begin{math}
\mu \end{math} and in the simple periodic
case only with respect to \begin{math}
\mu \end{math}. In fact, in both cases the
meromorphic function \begin{math} \lambda \end{math}
induces a finite covering of the Riemann
surface. Then there exists an unique
polynomial in \begin{math} \mu \end{math}
with coefficients depending
meromorphically on \begin{math}
\lambda \end{math}, which describes the
relation between \begin{math} \lambda \end{math}
and \begin{math} \mu \end{math} (see
\cite{Fo}). In the double periodic case
we will explain later how to define this
function \begin{math} R(\cdot ,\cdot)\end{math}. This relation
determines the Riemann surface \begin{math} \tilde Y \end{math}. In fact,
the transcendental curve \begin{math} \tilde Y \end{math} defined by the
equation \begin{math} R(\lambda ,\mu )=0\end{math}
may differ from the Riemann surface \begin{math}
Y\end{math} only by some
singularities\footnote{We assume, that
at all points \begin{math} y_{1},\ldots
,y_{n}\end{math} the greatest common divisor
of the degrees of the two fixed functions
in \begin{math} {\bf h}^{n}\end{math}
is equal to 1. Otherwise the Riemann
surface \begin{math} Y\end{math} may be
a non-trivial covering over the curve
defined by the equation \begin{math}
R(\lambda ,\mu )=0\end{math}. }.
If \begin{math} n\end{math} is greater
than 2, it may be convenient to modify our construction
and identify the points \begin{math}
y_{1},\ldots ,y_{n}\end{math} to one
multiple point. The corresponding singular Riemann
surface is denoted by \begin{math}
Y'\end{math}\footnote{The definition of
the Picard group of a singular Riemann
surface is given in \cite{Se}.}. An analogous construction
defines a group homomorphism from \begin{math}
H^n\end{math} into the Picard group of the
singular Riemann surface \begin{math}
Y'\end{math}.

}
Let us now give our main examples:
\begin{enumerate}
\item The completely integrable systems, investigated in the papers \cite{AvM}
and \cite{RST} are examples for the finite dimensional case.
\item The KdV equation with periodic boundary conditions is
obtained in the simple periodic case with \begin{math} n=1\end{math}. The
fixed elements are $f(z)=z$ and $f(z)=z^2$, $z^2$ generates the trivial flow
and $z$ generates the periodic spatial one (see e.g. \cite{Mu}).
\item The non-linear Schr{\"o}dinger equation with periodic boundary
conditions is obtained in the modified simple periodic case with \begin{math}
n=2\end{math}.  The fixed flows are  \begin{math} (z_{1},z_{2})\mapsto
(z_{1},z_{2})\end{math} (the trivial one) and
\begin{math} (z_{1},z_{2})\mapsto (\sqrt{-1}z_{1},-\sqrt{-1}z_{2})\end{math}
(the periodic one) (see \cite{HSS} and \cite{Sch}).
\item The KP equation with periodic boundary
conditions with respect to both space variables is obtained in the double
periodic case with \begin{math} n=1\end{math}
and the two functions corresponding to periodic flows are given by
\begin{math} z\mapsto z\end{math} and \begin{math} z\mapsto
\sigma z^{2}\end{math} (see \cite{Kr2}).
\end{enumerate}
The subject of this article is to consider the space of all data, which obey
the conditions introduced in the three cases above.  The methods of this paper
can be naturally generalized to infinite genus Riemann surfaces
with appropriate analytic properties (see e.g. \cite{Sch}).
Two data \begin{math} (Y,y_{1},\ldots ,y_{n},z_{1},\ldots
,z_{n})\end{math} and \begin{math}
(\tilde{Y},\tilde{y} _{1},\ldots ,\tilde{y} _{n},\tilde{z} _{1},\ldots
,\tilde{z} _{n})\end{math} are called
equivalent, if they may be mapped biholomorphically onto each other. It is
quite obvious, that two equivalent data either both obey the above mentioned
conditions or none of them. In fact, we are only interested in the set of
equivalence classes of data, which obey one of the above mentioned
conditions.

\subsection{The finite dimensional case}
Let us now consider the  finite dimensional case. Then those data,
which obey the conditions introduced above are in one to one correspondence
with plane algebraic curves defined by the equation \begin{math}
R(\lambda ,\mu )=0\end{math}. The highest powers
of this polynomial with respect to \begin{math}
\lambda \end{math}  and \begin{math} \mu \end{math}
are determined by the two fixed elements
of \begin{math} {\bf h}^{n}\end{math}.
It turns out that the polynomials \begin{math}
R(\cdot ,\cdot )\end{math}, which
correspond to some data obeying the above
condition of the finite dimensional
case, are exactly of the form
\begin{displaymath}
R_{0}(\lambda ,\mu )+\sum_{m,l}
r_{l,m}\lambda ^{l}\mu ^{m}\mbox{, with \begin{math}
r_{l,m}\in {\Bbb C}\end{math}.} \end{displaymath}
Here the sum is taken over some finite
convex set of \begin{math} {\Bbb N}_{0}\times
{\Bbb N}_{0}\end{math}. In fact let us
assume that \begin{math} R(\lambda ,\mu ,t)\end{math}
is a onedimensional family of
polynomials corresponding to some data
obeying these conditions. Then
the expression
\begin{eqnarray}
\frac{\partial \lambda }{\partial t} d\mu
-\frac{\partial \mu }{\partial t} d\lambda =
{-\frac{\partial R(\lambda ,\mu ,t)}{\partial t}
}\left( {\frac{\partial R(\lambda ,\mu ,t)}{\partial \lambda }
} \right) ^{-1}d\mu = \nonumber
\\ ={\frac{\partial R(\lambda ,\mu ,t)}{\partial t}
}\left( {\frac{\partial R(\lambda ,\mu ,t)}{\partial \mu }
} \right) ^{-1}d\lambda \nonumber
\end{eqnarray}
is a regular
meromorphic 1-form of the (in some cases singular) algebraic curve
defined by the equation \begin{math}
R(\lambda ,\mu ,t)=0\end{math}. This formula
shows that this form is regular on the
domain, where \begin{math} \lambda \end{math}
and \begin{math} \mu \end{math}  are
finite. The order of the pole of our form in the point \begin{math}
y_j \end{math} is not greater than the minimal of the orders of
$\lambda$ and $\mu$ in this point. We shall call such forms weakly singular.
If we want to compare the different Riemann
surfaces corresponding to different
values of \begin{math} t\end{math}, we
may choose either \begin{math} \lambda \end{math} not to depend on
\begin{math} t\end{math}, such that \begin{math} \mu \end{math}
becomes a multivalued function depending
on \begin{math} \lambda \end{math} and \begin{math}
t\end{math} or we choose \begin{math} \mu \end{math} not to depend on
\begin{math} t\end{math}, such that \begin{math} \lambda \end{math}
becomes a multivalued function depending
on \begin{math} \mu \end{math}  and \begin{math}
t\end{math}. In the first case \begin{math}
\frac{\partial \mu }{\partial t} d\lambda \end{math}
and in the second case \begin{math} \frac{\partial
\lambda }{\partial t} d\mu \end{math} is a weakly singular 1-form of our
singular Riemann surface. This allows us to identify the
tangent space of all equivalence classes obeying the condition
in the modified finite dimensional case with the space of
weakly singular 1-forms of the corresponding
singular Riemann surface. In fact, the parameters
\begin{math} r_{l,m}\end{math} give a global parameterization of the set
of all data obeying the conditions of the modified finite dimensional case.
The corresponding weakly singular 1-forms are given by
\begin{eqnarray}
\frac{\partial \lambda }{\partial r_{l,m}} d\mu
-\frac{\partial \mu }{\partial r_{l,m}} d\lambda = \nonumber \\
={-\lambda ^{l}\mu ^{m}}
\left( {\frac{\partial R(\lambda ,\mu ,t)}{\partial \lambda }}
 \right) ^{-1}d\mu =
{\lambda ^{l}\mu ^{m}
}\left( {\frac{\partial R(\lambda ,\mu ,t)}{\partial \mu }
} \right) ^{-1}d\lambda \nonumber
\end{eqnarray}
\begin{Example} \label{E1}
Let \begin{math} n\end{math} be equal to
2 and the two fixed elements of \begin{math}
{\bf h}^{2}\end{math} be given by
\begin{displaymath} (z_{1},z_{2})\mapsto
(z_{1},z_{2})\mbox{ and } (z_{1},z_{2})\mapsto
(z_{1}^{d},-z_{2}^{d})\mbox{, respectively.}
\end{displaymath} Then the corresponding
polynomials are exactly of the form
\begin{displaymath}
R(\lambda ,\mu )=\mu ^{2}-\lambda
^{2d}-\sum_{l=0}^{d-1}
r_{l}\lambda ^{l}\mbox{, with \begin{math}
r_{l}\in {\Bbb C}\end{math}.}
\end{displaymath}
{\tolerance=2000 The weakly singular 1-forms of the plane curves
defined by the equation \begin{math}
R(\lambda ,\mu )=0\end{math} are given by

}
\begin{displaymath}
\frac{\partial \lambda }{\partial r_{l}}
d\mu -\frac{\partial \mu }{\partial
r_{l}} d\lambda = -\frac{\lambda
^{l}}{2\mu } d\lambda \mbox{ with }
l=0,\ldots ,d-1.
\end{displaymath}
\end{Example}

\subsection{The simple periodic case}
Now we consider the modified simple periodic case. It corresponds to
solutions of $1+1$ systems, which are periodic in one spatial variable.
Due to this condition the function \begin{math}
R(\cdot ,\cdot )\end{math} is of the form
\begin{displaymath} R(\lambda ,\mu )=\mu ^{M}+\sum_{m=0}^{M-1}
\mu ^{m}r_{m}(\lambda )\end{displaymath}
with some entire functions
\begin{math} r_{0}(\cdot ),\ldots , r_{M-1}(\cdot )\end{math} of finite order
which have essential singularities at \begin{math} \lambda =\infty \end{math}.
Hence the transcendental curve defined by the equation \begin{math}
R(\lambda ,\mu )=0\end{math} has infinite algebraic genus. Let us again
assume that \begin{math} R(\lambda ,\mu ,t)\end{math}
is a onedimensional family of entire functions corresponding to some data
obeying these conditions. The quasimomentum $p$ is the multivalued function
defined by $\frac{1}{\sqrt{-1}} \ln(\mu )$. Then the expression
\begin{displaymath}
\frac{\partial p }{\partial t} d\lambda
-\frac{\partial \lambda }{\partial t} dp =
\end{displaymath}
\begin{displaymath} =
\der{R(\lambda ,\mu ,t)}{t}
\left( {\frac{\partial R(\lambda ,\mu ,t)}{\partial \lambda }
}\right) ^{-1}dp =
\sqrt{-1} {\frac{\partial R(\lambda ,\mu ,t)}{\partial t}
}\left( {\mu \frac{\partial R(\lambda ,\mu ,t)}{\partial \mu }
}\right) ^{-1}d\lambda
\end{displaymath}
is a regular meromorphic 1-form on \begin{math} \tilde Y\end{math}.
If the deformation of $R$ corresponds to a continuous deformation of $Y$
then this differential is also meromorphic on \begin{math} Y(t) \end{math}.
This formula shows again that this form is regular on the
domain where \begin{math} \lambda \end{math}
and \begin{math} \mu \end{math}  are
finite. The order of the pole of our form in the point
\begin{math} y_j \end{math} is not greater than the minimal of the orders of
$\lambda$ and $p$ in this point. We shall call such forms weakly
singular. Thus any weakly meromorphic
differential \begin{math} \omega =\nu d\lambda
\end{math} of \begin{math} Y \end{math} generates an isoperiodic flow
defined by
\begin{equation}
\label{Krtype}
\der{p}{t}d\lambda-\der{\lambda}{t}dp=\omega
\end{equation}
(This representation is rather similar to the form in which the
Whitham equations for the KP equation were written \cite{Kr3}).
Let \begin{math} \lambda_k \end{math} be a simple branch point of our
covering \begin{math} Y \rightarrow {\Bbb {CP}}^1 \end{math}. Then from (\ref
{Krtype}) it follows:
\begin{equation}
\label{difftype}
\der{\lambda_k}{t}=-\frac{\omega(\lambda_k)}{dp(\lambda_k)}
\end{equation}
(Similar representation is widely used in the Whitham theory \cite{FFM}).
If all the branch point are simple we get a system of ODE's but with a
rather complicated right-hand side.
If we want to compare the different Riemann
surfaces corresponding to different values of \begin{math} t\end{math}, we
may choose either \begin{math} \lambda \end{math} not to depend on
\begin{math} t\end{math}, such that \begin{math} \mu \end{math}
becomes a multivalued function depending
on \begin{math} \lambda \end{math} and \begin{math}
t\end{math}, or we choose \begin{math} \mu \end{math}
not to depend on \begin{math}
t\end{math}, such that \begin{math} \lambda \end{math}
becomes a multivalued function depending
on \begin{math} \mu \end{math}  and \begin{math}
t\end{math}. In the first case \begin{math}
\frac{\partial p }{\partial t}
d\lambda \end{math} and in the second case \begin{math} \frac{\partial
\lambda }{\partial t} dp \end{math} is a
weakly singular 1-form of our singular Riemann
surface. Thus we arrive at the following conclusion (see \cite{Sch}):
The tangent space to all data corresponding to solutions which are periodic
in one variable with given period is in one-to-one
correspondence with the space of
weakly singular 1-forms of the singular Riemann surface
\begin{math} \tilde Y(t) \end{math}. If we also admit Riemann
surfaces of infinite genus, the set of all functions
\begin{math} R(\cdot ,\cdot )\end{math}, which
correspond to data obeying the condition of the
simple periodic case, is again an affine space.
But then we have to take care of the boundary
conditions near the points \begin{math} y_{1},\ldots
,y_{n}\end{math}. The normalization \begin{math} Y \end{math}
of our transcendental curve \begin{math} \tilde Y \end{math} (the surface
obtained by removing all the singularities) is now algebraic and
coincides with the algebraic curve which plays the role of the spectral data
in the finite-gap theory.
Consider the following meromorphic function on  \begin{math} Y \end{math}
\begin{displaymath}
\kappa =\frac{dp }{d\lambda } =\frac{1}{\mu }
\frac{d\mu }{d\lambda } =-\frac{1}{\mu }
{\frac{\partial R(\lambda ,\mu )}{\partial \lambda }}/
{ \frac{\partial R(\lambda ,\mu )}{\partial \lambda }}.
\end{displaymath}
We may describe  \begin{math} Y \end{math}  by some equation
of the form \begin{displaymath}
Q(\kappa ,\lambda )=0.\end{displaymath}
Let us remark that this representation differs from the standard one.
This function \begin{math} Q(\cdot ,\cdot )\end{math}
is again uniquely defined as a polynomial in \begin{math} \kappa
\end{math} with coefficients depending meromorphically on
\begin{math} \lambda \end{math}, which
is even a meromorphic function on \begin{math}
{\Bbb {CP}}^{1}\times {\Bbb {CP}}^{1}\end{math}, if the corresponding
Riemann surface has finite genus.
Let us now assume that some regular 1-form \begin{math}
\omega \end{math} is given
by \begin{displaymath}
\omega =\nu d\lambda =\nu \kappa ^{-1}dp
\end{displaymath} with some
meromorphic function \begin{math} \nu \end{math}
on our Riemann surface. Then we may
calculate the derivative of the function
\begin{math} Q(\cdot ,\cdot )\end{math}
with respect to the corresponding
flow-parameter \begin{math} t\end{math}.
In order to do this we choose \begin{math}
\lambda \end{math} to depend not on \begin{math}
t\end{math}, such that both functions \begin{math}
\kappa \end{math} and \begin{math} \mu \end{math}
are multivalued functions depending on \begin{math}
\lambda \end{math}  and \begin{math}
t\end{math}. By definition we have the
equation \begin{displaymath}
\frac{\partial p }{\partial t} =\nu .\end{displaymath}
Equation (\ref{Krtype}) takes the form:
\begin{equation}
\label{FFMform}
\frac{\partial ^{2}p }{\partial
t\partial \lambda } =
\frac{\partial \kappa}{\partial t} =
\frac{\partial \nu }{\partial \lambda }
\mbox{ or equivalently }
\der{dp}{t}=d\nu .\end{equation}
This representation is similar to the Flaschka-Forest-McLaughlin representation
\cite{FFM} of the Whitham equations. Equation (\ref{FFMform}) has been
derived in the theory of the nonintegrable
periodic perturbations of the soliton equations in \cite{EFMS}.
Similar equations for meromorphic \begin{math} \omega \end{math} (which may be
not isoperiodic) were written in \cite{Kr1} formula (7.68)).
Since we have chosen \begin{math} \lambda \end{math}
not to depend on \begin{math} t\end{math}
we may calculate with this formula the
derivative of \begin{math} Q\end{math}
with respect to \begin{math}
t\end{math}: \begin{displaymath}
\frac{\partial Q(\kappa ,\lambda ,t)}{\partial t} =
-{\frac{\partial \kappa }{\partial t} }{
\frac{\partial Q(\kappa ,\lambda )}{\partial \kappa } }
=-\frac{\partial \nu }{\partial \lambda } {
\frac{\partial Q(\kappa ,\lambda )}{\partial \kappa } }
.\end{displaymath}  If we now could find
\begin{description}
\item[(i)] some parameterization of a set
of functions \begin{math} Q(\cdot ,\cdot )\end{math},
which contains the
functions describing the data obeying the
conditions in the modified simple
periodic case and
\item[(ii)] all
meromorphic functions \begin{math}
\nu \end{math}, such that \begin{math}
\nu d\lambda \end{math} is a weakly singular 1-form of
the Riemann surface described by \begin{math}
Q(\kappa ,\lambda )=0\end{math},
\end{description}
then this equation defines some flows on the
space of parameters, which leaves
invariant the subset of those values of
the parameters, which correspond to data
obeying our condition. If
furthermore  we \begin{description}
\item[(iii)] find one function \begin{math}
Q(\cdot ,\cdot )\end{math}, which
corresponds to some data obeying our
conditions, \end{description}
then we may determine a whole family of such
functions \begin{math} Q(\cdot ,\cdot)\end{math}.

In the previous considerations $\omega$ was an
arbitrary weakly singular form. We now present
a natural set of parameters.
Consider a canonical basis of cycles \begin{math}
a_{1},\ldots ,a_{g},b_{1},\ldots ,b_{g}\end{math}, such that the
intersection number between \begin{math}
a_{i}\end{math}  and \begin{math} b_{i}\end{math}
is equal to 1 for all \begin{math}
i=1,\ldots ,g\end{math}  and vanishes on
all other combinations (see \cite{Fo}).
Then the first homology group is isomorphic
to the free abelian group \begin{displaymath}
{\Bbb Z}a_{1}\oplus \ldots \oplus {\Bbb
Z}a_{g}\oplus {\Bbb Z}b_{1}\oplus
\ldots \oplus {\Bbb Z}b_{g}.\end{displaymath}
Since the intersection form is a
non-degenerate antisymmetric \begin{math}
{\Bbb Z}\end{math}-valued form, we may associate
to the function \begin{math} \mu \end{math} an
element \begin{math} {\rm m}\end{math} of
the first homology group, such that the
integral of the 1-form \begin{math} dp \end{math}
is equal to \begin{math} 2\pi \end{math} times the
intersection form with \begin{math} {\rm m}\end{math}.
By changing the basis we
may achieve that \begin{displaymath}
\oint_{a_{i}} dp =0 \mbox{ for all \begin{math}
i=1,\ldots ,g\end{math}} \Leftrightarrow
{\rm m} \in {\Bbb Z}a_{1}\oplus \ldots \oplus
{\Bbb Z}a_{g}.\end{displaymath}
Then there exists a unique basis \begin{math}
\omega _{1},\ldots ,\omega _{g}\end{math} of holomorphic
1-forms of our compact Riemann
surface, such that
\begin{equation}
\oint_{a_{i}} \omega _{j}=2\pi \sqrt{-1} \delta _{i,j}\mbox{ for all }
i,j\in \{ 1, \ldots ,g\} .
\label{basis}\end{equation}
Now we can define \begin{math} g\end{math}
flows on the space of compact Riemann
surfaces of genus \begin{math}
g\end{math}, which obey the condition
of the simple periodic case:
\begin{equation}
\frac{\partial p }{\partial t_{i}}
d\lambda -\frac{\partial \lambda }{\partial t_{i}}
dp =\omega _{i}\mbox{ for all }
i=1,\ldots ,g.
\label{eq1}
\end{equation}
Of course the canonical basis extends
uniquely to some neighbourhood of each
of these Riemann surfaces. But globally
the space of Riemann surfaces together
with some canonical basis will be a
non-trivial covering over the space of Riemann surfaces.
It is quite easy to see, that the
parameters \begin{math} t_{1},\ldots
,t_{g}\end{math} locally may be chosen
to be given by \label{loctimes}
\begin{equation}
t_{i}=\frac{1}{2\pi \sqrt{-1}} \oint_{a_{i}} p d\lambda \mbox{ for all }
i=1,\ldots ,g.
\label{times-per}
\end{equation}
These parameters depend only on the
choice of the canonical basis obeying
the foregoing condition.
Moreover, this shows
that all these flows commute.
Now it is obvious that there is a number which
does not change under this flows:

\noindent
The greatest number \begin{math}
m\in {\Bbb N}\end{math}, such that \begin{math}
{\rm m}/m\end{math} is an element of the
first homology group. We will see in the example
of the nonlinear Schr{\"o}dinger equation
that this number classifies the connected components
of the set of data obeying the conditions of the
modified periodic case.

\begin{Remark}
These flows can be extended to Riemann surfaces corresponding to general
finite-gap solutions of soliton equations with one spatial dimension (in
general they are quasiperiodic). In this case the `isoperiodic deformations'
preserve the group of the frequences of these solutions.
So we consider all data which obey the following
condition in the
\begin{description}
\item[simple quasiperiodic case:] one of the fixed elements in
\begin{math} H^n\end{math} generates a trivial flow and no conditions
on the flow generated by the second fixed element are imposed.
\end{description}
The function
\begin{math} \lambda \end{math} corresponding to the trivial flow maps
\begin{math} Y \end{math} to \begin{math} {\Bbb {CP}}^1 \end{math} so
\begin{math} Y \end{math} is represented as a ramified covering of
\begin{math} {\Bbb {CP}}^1 \end{math}. The position of the branch points
gives us local coordinates on the space of such data.

In this case an analogue of the quasimomentum differential
can be defined uniquely as an Abel differential \begin{math} dp \end{math}
of the second kind such that:

\begin{description}
\item[diff 1:] \begin{math} dp \end{math} is holomorphic in
\begin{math} Y\setminus (y_1\cup y_2 \cup \ldots \cup y_n) \end{math}.
\item[diff 2:] \begin{math}
dp=\frac1{\sqrt{-1}}df_k+\hbox{reg.terms} \end{math}
in the point
\begin{math} y_k \end{math} where \begin{math} (f_1,\ldots,f_n) \end{math}
is the second fixed element of \begin{math} H^n \end{math}
\item[diff 3:] \begin{math} \hbox{Im} \oint_c dp=0\end{math} for any closed
cycle
\begin{math} c \end{math}.
\end{description}

\noindent
\begin{math} \kappa=dp/d\lambda \end{math} is again a meromorphic
function on \begin{math} Y \end{math} and the foregoing discussion
of the isoperiodic flows carries over to this more general situation.
In fact, the right-hand side of (\ref{FFMform})
is an exact differential so all
the periods of the left-hand side are equal to $0$. Thus all the periods
of \begin{math} dp \end{math} are integrals of motion.
\end{Remark}
Let us consider concrete equations in more detail.

\begin{Example} \label{E2}
As we mentioned above, the KdV equation with periodic boundary
conditions is an example of the simple periodic case \cite {N1}.
It is well known (see e.g.\cite{MKM}), that in
this case the function \begin{math}
R(\cdot ,\cdot )\end{math} is of the form
\begin{displaymath}
R(\lambda ,\mu )=\mu ^{2}-\mu \Delta (\lambda )+1.
\end{displaymath}
Hence we may calculate the function
\begin{math} \kappa \end{math} in terms of \begin{math} \lambda \end{math}
and \begin{math} \mu \end{math}:
\begin{displaymath}
\kappa =\frac{\sqrt{-1} }{\mu } \frac{\partial R(\lambda ,\mu )}{\partial
\lambda } /\frac{\partial R(\lambda ,\mu )}{\partial
\mu } =-\frac{\partial \Delta (\lambda )}{\partial \lambda }
\frac{\sqrt{-1} }{2\mu -\Delta (\lambda )} .
\end{displaymath}
Hence the function \begin{math} Q(\cdot ,\cdot ) \end{math} has the form
\begin{displaymath}
Q(\kappa ,\lambda )=\kappa ^{2}-\left( \frac{\partial
\Delta (\lambda )}{\partial \lambda } \right )^{2}
\frac{1}{\Delta ^{2}(\lambda )-4} .
\end{displaymath}
If the Riemann surface has finite genus, there
are only a finite number of zeroes of \begin{math}
\frac{\partial \Delta (\lambda )}{\partial \lambda }
\end{math}, where the denominator \begin{math}
\Delta ^{2}(\lambda )-4\end{math} is not zero too.
Moreover, by definition of the two fixed
elements of \begin{math} {\bf
h}\end{math}: to the function \begin{math}
\lambda \end{math} there corresponds the
element \begin{math} z\mapsto z^{2}\end{math}
and to the function \begin{math} \ln \mu \end{math}
there corresponds the function \begin{math}
z\mapsto z\end{math}, hence for very large \begin{math}
\lambda \end{math} the function \begin{math}
\kappa ^{2}\end{math} is almost equal to \begin{math}
1/4\lambda \end{math}. Then \begin{math}
Q(\cdot ,\cdot )\end{math} is of the
form \begin{displaymath}
Q(\kappa ,\lambda )=\kappa ^{2}-\frac{\prod_{i=1}^{g}
(\lambda -\alpha _{i})^{2}}{4\prod_{j=0}^{2g}(\lambda -\lambda _{j})},
\end{displaymath}
where \begin{math} g\end{math}
is the genus, \begin{math}
\alpha _{1},\ldots ,\alpha _{g}\end{math} are the
values of the function \begin{math}
\lambda \end{math} at the \begin{math} 2g\end{math}
zeros of the differential \begin{math}
dp \end{math} and \begin{math}
\lambda _{0},\ldots ,\lambda _{2g},\infty \end{math}
are the values of the function \begin{math}
\lambda \end{math} at the zeroes of the
differential \begin{math} d\lambda \end{math}
(these zeroes are of course the
branchpoints of the covering map induced
by \begin{math} \lambda \end{math}). The standard representation
for this Riemann surface differs from this one and reads as:
\begin{equation}
\label{kdv-sur}
\zeta ^2=4 (\lambda-\lambda_0)\ldots(\lambda-\lambda_{2g}).
\end{equation}
Now the holomorphic 1-forms of the Riemann
surface defined by the equation \begin{math}
Q(\kappa ,\lambda )=0\end{math} are of the form
\begin{equation}
\omega =\frac{o(\lambda ) d\lambda}
{2 \sqrt{d(\lambda)}} \label{kdv-form}
\end{equation}
Here \begin{math} o(\cdot )\end{math}
denotes a function \begin{math}
\sum_{i=0}^{g-1} o_{i}\lambda ^{i}\end{math},
\begin{math} d(\cdot )\end{math} denotes the
function \begin{math}
d(\lambda )=\prod_{j=0}^{2g} (\lambda -\lambda _{j})\end{math}
and \begin{math} q(\cdot )\end{math}
denotes the function \begin{math}
q(\lambda )=\prod_{i=1}^{g}
(\lambda -\alpha _{i})\end{math}.
The flow corresponding to this
1-form is given by the differential
equation
\begin{equation}
\label{kdv-diffeq}
2d(\lambda )\frac{\partial q(\lambda )}{\partial t}
-q(\lambda )\frac{\partial d(\lambda )}{\partial t}
=\left ( 2d(\lambda )\frac{\partial o(\lambda )}{\partial \lambda }
-o(\lambda )\frac{\partial d(\lambda )}{\partial \lambda } \right ).
\end{equation}
This flow may also be formulated in terms of \begin{math}
\Delta (\cdot )\end{math}:
\begin{displaymath} \frac{\partial \Delta (\lambda )}{\partial t}
=\frac{o(\lambda )}{q(\lambda )}
\frac{\partial \Delta (\lambda )}{\partial \lambda } .
\end{displaymath}
The flow is written as a system of ordinary
differential equations (\ref{kdv-diffeq}),
but this system is rather complicated,
because the coefficients of the polynomial \begin{math}q(\lambda) \end{math}
(or, equivalently, the zeroes of \begin{math}q(\lambda) \end{math}) depend
on the branch points in a rather complicated way (they can be expressed
in terms of some hyperelliptic integrals). But we may consider
$\lambda_0$, \ldots, $\lambda_{2g}$, $\alpha_0$, \ldots, $\alpha_g$,
as independent variables. This means that the differential
\begin{equation}
\label{kdv-dp}
dp=\frac{q(\lambda) }{2 \sqrt{d(\lambda)}} d \lambda=\kappa d\lambda.
\end{equation}
will have arbitrary periods. Neverthelesson this bigger set of parameters
(\ref{kdv-diffeq}) still defines an ordinary differential
equation.
%\begin{eqnarray}
%\der{\lambda_k}{t}=-\frac{o(\lambda_k)}{q(\lambda_k)},
%\nonumber
%\\
%\label{kdv-req}
%\der{\alpha_k}{t}=\frac{-1}
%{\prod\limits_{j\ne k}(\alpha_k-\alpha_j)}
%\left[o'(\alpha_k) -\frac12 \left(\sum\limits_{j=0}^{2g}
%\frac{1}{\alpha_k-\lambda_j} \right ) o(\alpha_k) \right ].
%\end{eqnarray}
Moreover, the coefficients of $o(\cdot )$ may depend
on all these parameters in a rather complicated way, which is the case
in the choice of (\ref{basis}) and (\ref{eq1}).
So let us present another choice, in which (\ref{kdv-diffeq})
simplifies to ODE's with a rational right-hand side.
If all the values \begin{math} \alpha _{1},\ldots
,\alpha _{g},\lambda _{0},\ldots ,\lambda _{2g}\end{math}
are pairwise different, we have a basis of holomorphic 1-forms
\begin{displaymath}
\omega _{i}=-\frac{c_{i}}{\lambda
-\alpha _{i}} dp \mbox{ for all }
i=1,\ldots ,g\mbox{ with some } c_{i}\in
{\Bbb C}\setminus \{ 0\} .
\end{displaymath}
In the periodic case the foregoing formula shows, that the
values of the function \begin{math}
\Delta (\alpha _{i})\end{math} depend only on
the flow parameter corresponding to the
form \begin{math} \omega _{i}\end{math}. It
is well know, that \begin{math}
\Delta (\lambda )=2\cos \left( \sqrt{\lambda
-\lambda _{0}} \right)
\end{math}
describes the Riemann surface (of genus
0) corresponding to constant
potentials. This transcendental curve has
infinitely many double points at the
values \begin{math} \alpha _{i}=(\lambda _{0}+i\pi
)^{2}\end{math} for all \begin{math}
i\in {\Bbb N}\end{math}. So let us choose
this curve as the starting point of our
flows and normalize the 1-forms by the
condition \begin{equation}
\label{norm}
\Delta (\alpha _{i})=(-1)^{i}2\cosh (t_{i})\mbox{ for all }
i\in {\Bbb N}.\end{equation}
So we introduce an infinite sequence of flow
parameters. But we are interested in the
case, where only finitely many  of these parameters
are not equal to zero. This is precisely
the case of Riemann surfaces of finite genus. In fact,
generically the genus is equal to the number
of flow parameters, which are not equal to zero.
The corresponding differential equations
are given by \begin{equation}
\label{rational1}
\frac{\partial \lambda _{j}}{\partial t_{i}}
=-\frac{c_{i}}{\lambda _{j}-\alpha _{i}} \mbox{ with }
j=0,\ldots ,2g;
\end{equation}

\begin{eqnarray}
\frac{\partial \alpha _{j}}{\partial t_{i}}
=-\frac{c_{i}}{\alpha _{j}-\alpha _{i}} \mbox{ with }
j=1,\ldots,i-1,i+1,\ldots ,g; \nonumber \\
\frac{\partial \alpha _{i}}{\partial t_{i}}
=\sum_{j\neq i}
\frac{c_{i}}{\alpha _{j}-\alpha _{i}}
-\frac{1}{2} \sum_{j=0}^{2g}
\frac{c_{i}}{\lambda _{j}-\alpha _{i}}.
\label{rational2}
\end{eqnarray}
The normalization condition implies
\begin{displaymath}
c_{i}^{2}=\frac{\prod_{j=0}^{2g}(\alpha _{i}-\lambda _{j})}{
\prod_{j\neq i}
(\alpha _{i}-\alpha _{j})^{2}}.\end{displaymath}
Locally this uniquely defines the coefficients \begin{math}
c_{1},\ldots ,c_{g}\end{math}. Due to
the normalization condition all these
flows commute locally.
It is quite easy to see that the
singularity of these flows at the
starting point may be removed. The \begin{math}
i\end{math}-th flow opens the \begin{math}
i\end{math}-th gap. For real flow
parameters these flows have no
singularities, but in the complex domain
they have very complicated
singularities. The forgoing formula implies
that \begin{math} \sum_{j=0}^{2g}
\lambda _{j}-2\sum_{i=1}^{g} \alpha _{i}\end{math}
does not depend on \begin{math}
t\end{math}. It is well known, that this constant
may be an arbitrary number in \begin{math}
{\Bbb C}\end{math} (it is related to the mean value
of the periodic potential). The integrals over all
generators of the first homology group of the form
\begin{math} dp \end{math} are also
invariant under these flows. Hence we have \begin{math}
2g+1\end{math} integrals of motion.
Our space of parameters has dimension
\begin{math} 3g+1\end{math}. This shows that
the set of data with genus \begin{math} g\end{math}
obeying the conditions of the
periodic KdV equation has dimension
\begin{math} g+1\end{math}.

Consider now the general finite-gap quasiperiodic KdV solutions. The formulas
(\ref{kdv-sur}), (\ref{kdv-diffeq}), (\ref{rational1}) and (\ref{rational2})
are valid in this case too. But in the quasiperiodic
case we have no natural analogues of \begin{math}
\Delta(\lambda)\end{math}. Nevertheless, the
function \begin{math}p(\lambda)\end{math}
is locally well-defined and the normalization condition
(\ref{norm}) is solved by the local parameters
\begin{math}\sqrt{-1}p(\alpha_i)+i\pi \end{math},
\begin{math}i=1,\ldots,g\end{math} of the moduli
space of Riemann surfaces corresponding to the solutions with a given set
of frequencies. Similar coordinates were used by Marchenko \cite{Mar} in the
periodic KdV theory and by Krichever in the general situation (private
communication). All the periods of $dp$
are integrals of motion of (\ref{kdv-diffeq}). In the
space of all $\lambda_k$,  $\alpha_k$, we have a subspace defined by:
\begin{displaymath}
\hbox{Im} \oint dp= 0
\end{displaymath}
and the flows (\ref{kdv-diffeq}) with arbitrary \begin{math} o(\lambda)
\end{math}
leave this subspace invariant. Thus it is sufficient for us to know one point
of this subspace to construct the whole family. In fact these conditions are
integrals of motion of our equations.
\end{Example}
\begin{Remark}
H. Babujian has observed that the right-hand side of (\ref{rational2}) with
$c_i=1$ appeared also in the quasiclassical limit of the Bethe ansatz
equations, which are the Bethe ansatz equations for the Gaudin
magnet \cite{Ba}. The roots of this similarity are not clear now.
\end{Remark}
\begin{Example} \label{E3}
Let us consider now the non-linear
Schr{\"o}dinger equation as a second example
of the simple periodic case. It is very
similar to the KdV
equation. So let us only mention the
modifications.
The function \begin{math}
R(\cdot ,\cdot )\end{math} is again of the
form \begin{displaymath}
R(\lambda ,\mu )=\mu ^{2}-\mu \Delta
(\lambda )+1.\end{displaymath}
The corresponding Riemann surfaces of
finite genus have two covering points of
\begin{math} \lambda =\infty \end{math}, which
are identified to one multiple point.
They are described again by the equation
\begin{displaymath}
Q(\kappa ,\lambda )=\kappa ^{2}-\frac{\prod_{i=1}^{g}
(\lambda -\alpha _{i})^{2}}{
\prod_{j=1}^{2g}(\lambda -\lambda _{j})}
=0,\end{displaymath}
where \begin{math} g\end{math} is the algebraic genus of this singular
Riemann surface (this is equal to the
genus of the normalization plus one), \begin{math}
\alpha _{1},\ldots ,\alpha _{g}\end{math} are the
values of the function \begin{math}
\lambda \end{math} at the \begin{math} 2g\end{math}
zeros of the differential \begin{math}
dp \end{math} and \begin{math}
\lambda _{1},\ldots ,\lambda _{2g}\end{math}
are the values of the function \begin{math}
\lambda \end{math} at the zeroes of the
differential \begin{math} d\lambda \end{math}
(these zeroes are of course the
branchpoints of the covering map induced
by \begin{math} \lambda \end{math}).

The standard representation for this Riemann surface has the form
\begin{displaymath}
\zeta ^2=(\lambda-\lambda_1) \ldots (\lambda-\lambda_{2g})=d(\lambda ).
\end{displaymath}

Now the regular holomorphic 1-forms of the
singular Riemann
surface defined by the equation \begin{math}
Q(\kappa ,\lambda )=0\end{math} are of the form
\begin{displaymath}
\omega =\sum_{i=0}^{g-1} \frac{o_{i}\lambda ^{i}}{\sqrt {d(\lambda)}}
d\lambda \mbox{ with }
o_{i}\in {\Bbb C}\mbox{ for all }
i=0,\ldots ,g-1.
\end{displaymath}
\begin{displaymath}
dp=\frac{q(\lambda)}{\sqrt {d(\lambda)}} d\lambda.
\end{displaymath}

If all
the values \begin{math} \alpha _{1},\ldots
,\alpha _{g},\lambda _{1},\ldots ,\lambda _{2g}\end{math}
are pairwise different, we have again a
basis of regular 1-forms \begin{displaymath}
\omega _{i}=-\frac{c_{i}}{\lambda -\alpha _{i}}
dp \mbox{ for all }
i=1,\ldots g\mbox{ with some } c_{i}\in
{\Bbb C}\setminus \{ 0\} .\end{displaymath}
The values of the function \begin{math}
\Delta (\alpha _{i})\end{math} again only depend on
the \begin{math} i\end{math}-th flow
parameter. The function \begin{math}
\Delta (\lambda )=2\cos (\lambda )\end{math}
describes the Riemann surface (of genus
0) corresponding to the zero
potentials. So let us choose
this curve as the starting point of our
flows and normalize the 1-forms by the
condition \begin{displaymath}
\Delta (\alpha _{i})=(-1)^{i}2\cosh (t_{i})\mbox{ for all }
i\in {\Bbb Z}.\end{displaymath}
Again we introduce an infinite sequence of flow
parameters. But mostly we assume all but
a finite number of parameters to be
equal to zero. This is again
the case of Riemann surfaces of finite genus. In fact,
generically the algebraic genus is equal to the number
of flow parameters, which are not equal to zero.
The corresponding differential equations
are given by
\begin{displaymath}
\frac{\partial \lambda _{j}}{\partial t_{i}}
=-\frac{c_{i}}{\lambda _{j}-\alpha _{i}} \mbox{ with }
j=1,\ldots ,2g;
\end{displaymath}
\begin{displaymath}
\frac{\partial \alpha _{j}}{\partial t_{i}}
=-\frac{c_{i}}{\alpha _{j}-\alpha _{i}} \mbox{ with }
j=1,\ldots ,i-1,i+1,\ldots g;
\end{displaymath}
\begin{displaymath}
\frac{\partial \alpha _{i}}{\partial t_{i}}
=\sum_{j\neq i}
\frac{c_{i}}{\alpha _{j}-\alpha _{i}}
-\frac{1}{2} \sum_{j=1}^{2g}
\frac{c_{i}}{\lambda _{j}-\alpha _{i}} .
\end{displaymath}
The normalization condition implies
\begin{displaymath}
c_{i}^{2}=\frac{\prod_{j=1}^{2g}(\alpha _{i}-\lambda _{j})}{
\prod_{j\neq i} (\alpha _{i}-\alpha _{j})^{2}}.\end{displaymath}
Due to the normalization condition all these
flows commute locally.
It is quite easy to see that the
singularities of these flows at the
starting point may be removed. The \begin{math}
i\end{math}-th flow opens the \begin{math}
i\end{math}-th gap. For real flow
parameters these flows have again no
singularities, but in the complex domain
they have very complicated
singularities.
The forgoing formula implies
that \begin{math} \sum_{j=1}^{2g}
\lambda _{j}-2\sum_{i=1}^{g} \alpha _{i}\end{math}
does not depend on \begin{math}
t\end{math}. Due to the condition that near \begin{math}
\lambda =\infty \end{math} \begin{math} p
\end{math} has near \begin{math} \lambda =\infty
\end{math} two branches of the form \begin{math}
\pm \lambda +{\bf O}(\lambda ^{-1})
\end{math} this constant must be equal
to zero and furthermore, the integral of the 1-form
\begin{math} dp \end{math} along a path connecting
the two covering points of \begin{math}
\lambda =\lambda _{0} \end{math} must be of the form
\begin{math} \pm (2\lambda _{0}+2m\pi )+
\mbox{\bf O}(\lambda _{0}^{-1})\end{math}
near the point \begin{math} \lambda _{0}=\infty
\end{math}.
Moreover, the integrals over \begin{math} 2g-2\end{math}
generators of the first homology group of the form
\begin{math} dp \end{math} are also
invariant under these flows. Hence we have \begin{math}
2g\end{math} integrals of motion.
Our space of parameters has dimension
\begin{math} 3g\end{math}. This shows that
the set of data with algebraic genus \begin{math} g\end{math}
obeying the conditions of the
periodic nonlinear Schr{\"o}dinger equation has dimension
\begin{math} g\end{math}.
\end{Example}
\subsection{The double periodic case}
Let us now consider the double periodic
case. Our main example will be the KP equation (see \cite{Kr2}).
In general the function \begin{math} R(\cdot
,\cdot )\end{math} , which describes the
relation between the functions \begin{math}
\lambda \end{math}  and \begin{math}
\mu \end{math}, is not (uniquely) defined.
We assume, that for all \begin{math}
\lambda \in {\Bbb C}\setminus \{ 0\} \end{math}
the sum over the sheets of the covering
map induced by \begin{math} \lambda \end{math}
of the function \begin{math} |\mu |^{-1}\end{math}
converges. In this case we may define
the function \begin{math} R(\lambda ,\mu )\end{math}
in the following manner: \begin{displaymath}
R(\lambda ,\mu )=\prod_{\mbox{{\scriptsize over
the sheets of the covering map induced
by \begin{math} \lambda \end{math} }} } \left(
1-\frac{\mu }{\mu (\lambda )} \right) .\end{displaymath}
This choice may be characterized by two
properties: \begin{description}
\item[(i)] For all fixed \begin{math}
\lambda _{0}\in {\Bbb C}\setminus \{ 0\} \ \
R(\lambda _{0},\cdot )\end{math} is an entire
function with genus equal to zero.
\item[(ii)] For all \begin{math}
\lambda \in {\Bbb C}\setminus \{ 0\} \ \
R(\lambda ,0)=1\end{math}.
\end{description}
Now let us again assume that \begin{math}
R(\lambda ,\mu ,t)\end{math} is a one
dimensional family of such functions,
corresponding to some data obeying the
condition of the double periodic case.
Then the expression \begin{displaymath}
\frac{\partial \ln \lambda }{\partial t}
d\ln \mu
-\frac{\partial \ln \mu }{\partial t}
d\ln \lambda =
\end{displaymath}
\begin{displaymath}
=-\frac{\partial R(\lambda ,\mu ,t)}{\partial t}
\left( \lambda {\frac{\partial
R(\lambda ,\mu ,t)}{\partial \lambda }
}\right) ^{-1}d\ln \mu =
\end{displaymath}
\begin{displaymath}
={\frac{\partial R(\lambda ,\mu ,t)}{\partial t}
}\left( {\mu \frac{\partial R(\lambda ,\mu ,t)}{\partial \mu }
}\right) ^{-1}d\ln \lambda \end{displaymath}
is a regular
meromorphic 1-form of the (in some cases
singular) algebraic curve defined by the
equation \begin{math}
R(\lambda ,\mu ,t)=0\end{math}. This
formula shows again that this form is
regular on the domain, where \begin{math}
\lambda \end{math}  and \begin{math} \mu \end{math}
 are finite. Due to the conditions on
the functions \begin{math} \lambda \end{math}
and \begin{math} \mu \end{math} this form
has poles of order at most one at all
the points \begin{math} y_{1},\ldots
y_{n}\end{math}. Hence it is a regular
form of the singular Riemann surface,
where all the points \begin{math}
y_{1},\ldots ,y_{n}\end{math} are
identified to one multiple point.
If we want to compare the different Riemann
surfaces corresponding to different
values of \begin{math} t\end{math}, we
may choose either \begin{math} \lambda \end{math}
not to depend on \begin{math}
t\end{math}, such that \begin{math} \mu \end{math}
becomes a multivalued function depending
on \begin{math} \lambda \end{math} and \begin{math}
t\end{math} or we choose \begin{math} \mu \end{math}
not to depend on \begin{math}
t\end{math}, such that \begin{math} \lambda \end{math}
becomes a multivalued function depending
on \begin{math} \mu \end{math}  and \begin{math}
t\end{math}. In the first case \begin{math}
\frac{\partial \ln \mu }{\partial t}
d\ln \lambda \end{math}
and in the second case \begin{math} \frac{\partial
\ln \lambda }{\partial t} d\ln \mu \end{math} is a
regular 1-form of our singular Riemann
surface. This again allows us to identify the
tangent space of all equivalence classes
obeying the condition in the modified finite
dimensional case with the space of
regular 1-forms of the corresponding
singular Riemann surface. Let us now
restrict to the case \begin{math}
n=1\end{math}. In this case the Riemann
surface \begin{math} Y\end{math} may be
chosen to be the non-singular
normalization of the transcendental
curve described by the equation \begin{math}
R(\lambda ,\mu ,t)=0\end{math}. Moreover, we
restrict ourselves to the case in which
this normalization is a compact Riemann
surface of finite genus \begin{math}
g\end{math}. Then there exists a
canonical basis of cycles \begin{math}
a_{1},\ldots ,a_{g},b_{1},\ldots
,b_{g}\end{math}, such that the
intersection number of \begin{math}
a_{i}\end{math}  and \begin{math} b_{i}\end{math}
is equal to 1 for all \begin{math}
i=1,\ldots ,g\end{math}  and vanishes on
all other combinations (see \cite{Fo}).
Then the first homology group is isomorphic
to the free abelian group \begin{displaymath}
{\Bbb Z}a_{1}\oplus \ldots \oplus {\Bbb
Z}a_{g}\oplus {\Bbb Z}b_{1}\oplus
\ldots \oplus {\Bbb Z}b_{g}.\end{displaymath}
Since the intersection form is a
non-degenerate antisymmetric \begin{math}
{\Bbb Z}\end{math}-valued form, to both
functions \begin{math} \lambda \end{math} and \begin{math}
\mu \end{math} we may associate
elements \begin{math} {\rm l}\end{math}
and \begin{math} {\rm m}\end{math} of
the first homology group, such that the
integral of the 1-forms \begin{math}
d\ln \lambda \end{math} and \begin{math} d\ln
\mu \end{math} is equal to \begin{math}
2\pi \sqrt{-1}\end{math} times the
intersection form with \begin{math} {\rm
l}\end{math}  and \begin{math} {\rm m}\end{math}
respectively. By changing the basis we
may always achieve that either \begin{description}
\item[(i)] \begin{math}
\oint_{a_{i}} d\ln \lambda =0\end{math} for all \begin{math}
i=1,\ldots ,g\end{math},

\noindent or
\item[(ii)] \begin{math}
\oint_{a_{i}} d\ln \mu =0\end{math} for all \begin{math}
i=1,\ldots ,g\end{math}. \end{description}
Furthermore, if the intersection number of
\begin{math} {\rm l}\end{math}  and \begin{math}
{\rm m}\end{math}  is equal to zero, we
may even attain, that both conditions
are fulfilled.
{\tolerance=2000 \begin{Lemma}
If there exists a double periodic non-singular
solution corresponding to the compact Riemann surface
\begin{math} Y \end{math}, then the intersection
number of \begin{math} {\rm l}\end{math} and
\begin{math} {\rm m}\end{math} is zero.
\end{Lemma}

}
Proof: Both elements \begin{math} {\rm l}\end{math} and
\begin{math} {\rm m}\end{math} can be considered
as elements of the Lie algebra of the Picard group
of \begin{math} Y\end{math}. Now it is well known
that any double periodic solution corresponding
to the compact Riemann surface \begin{math} Y\end{math}
may be described by the family of divisors
\begin{displaymath}
D(s,t)=\exp (s{\rm l}+t{\rm m})D\mbox{ with some divisor
\begin{math} D\end{math} of degree \begin{math} g\end{math}
and } (s,t)\in {\Bbb R}^{2}.
\end{displaymath}
If this solution is non-singular, all divisors of this
family are non-special and have support inside of
\begin{math} Y\setminus \{ y_{1},\ldots y_{n}\} \end{math}.
Hence all these divisors may
be described by an unique \begin{math} g\end{math}-tuple
of points of \begin{math} Y\end{math}. Since both flows
are assumed to be periodic, the two parameters
\begin{math} (s,t)\end{math} of the family run through
\begin{math} ({\Bbb R}/{\Bbb Z})^2\end{math}.
The summation over the points of the divisors
defines a map \begin{displaymath}
H_{*}(({\Bbb R}/{\Bbb Z})^2,{\Bbb Z}) \rightarrow
H_{*}(Y,{\Bbb Z}).\end{displaymath}
Under this map the two generators of \begin{math}
H_{1}(({\Bbb R}/{\Bbb Z})^2,{\Bbb Z})\end{math}
are mapped onto \begin{math} {\rm l}\end{math}
and \begin{math} {\rm m}\end{math}. Moreover, this map
respects the intersection number. If the intersection
number of \begin{math} {\rm l}\end{math}
and \begin{math} {\rm m}\end{math} is not zero,
the image of the generator of \begin{math}
H_{2}(({\Bbb R}/{\Bbb Z})^2,{\Bbb Z})\end{math}
cannot be equal to zero. Then the union of the
paths of all points of the divisors of the whole
family covers the whole Riemann surface \begin{math}
Y\end{math}. This is a contradiction
to the assumption, that all divisors of the whole
family are non-special and have support inside of
\begin{math} Y\setminus \{ y_{1},\ldots y_{n}\} \end{math}.
\hspace*{\fill }\begin{math} \Box \end{math}

\noindent
In the sequel we assume both
conditions (i) and (ii).
Then there exists a unique basis \begin{math}
\omega _{1},\ldots ,\omega _{g}\end{math} of holomorphic
1-forms of our compact Riemann
surface, such that \begin{displaymath}
\oint_{a_{i}} \omega _{j}=\delta _{i,j}\mbox{ for all }
i,j\in \{ 1, \ldots ,g\} .\end{displaymath}
Now we can define \begin{math} g\end{math}
flows on the space of compact Riemann
surfaces of genus \begin{math}
g\end{math}, which obey the condition
of the double periodic case: \begin{displaymath}
\frac{\partial \ln \lambda }{\partial t_{i}}
d\ln \mu
-\frac{\partial \ln \mu }{\partial t_{i}}
d\ln \lambda =\omega _{i}\mbox{ for all }
i=1,\ldots ,g.\end{displaymath}
The canonical basis of course extends
uniquely to some neighbourhood of each
of these Riemann surfaces. But globally
the space of Riemann surfaces together
with some canonical basis will be a
non-trivial covering over the space of
Riemann surfaces.
It is quite easy to see, that locally the
parameters \begin{math} t_{1},\ldots
,t_{g}\end{math}  may be chosen
to be given by \begin{displaymath}
t_{i}=\oint_{a_{i}} \ln (\mu ) d\ln (\lambda ) \mbox{ for all }
i=1,\ldots ,g.\end{displaymath} Due to
our assumption the multivalued function \begin{math}
\ln \mu \end{math} may be
chosen to be single valued on each of the
cycles \begin{math} a_{1},\ldots
,a_{g}\end{math}. Hence these parameters
are locally defined only up to summation
of some elements of \begin{math} {\Bbb
Z}2\pi \sqrt{-1}\end{math}. If the intersection
number of \begin{math} {\rm l}\end{math}
and \begin{math} {\rm m}\end{math}  is
zero, these parameters depend only on the
choice of the canonical basis obeying
both conditions (i) and (ii).
Now it is obvious that three numbers
do not change under these flows:
\begin{enumerate}
\item the greatest number \begin{math}
l\in {\Bbb N}\end{math}, such that \begin{math}
{\rm l}/l\end{math} is an element of the
first homology group.
\item the greatest number \begin{math}
m\in {\Bbb N}\end{math}, such that \begin{math}
{\rm m}/m\end{math} is an element of the
first homology group too, and
\item the intersection number between \begin{math}
{\rm l}\end{math}  and \begin{math} {\rm m}\end{math}.
\end{enumerate}

Isoperiodic flows can be naturally extended to the general quasiperiodic
finite-gap solutions. Let $Y$ be a compact Riemann surface. Then any
holomorphic differential on $Y$ generates a deformation preserving all
the $x$ and $y$ frequences.

\section{Gradient flows in the simple periodic case}

In the section `simple periodic case' the isoperiodic flows were written in
terms of the branch points of $Y$ $\lambda_1$, \ldots, $\lambda_N$ as a
system of ordinary differential equations. These equations are defined for the
finite-gap quasiperiodic solutions as well as for the periodic ones.
In this case these flows are defined on the space of nonsingular compact
Riemann surfaces such that one of the fixed elements is represented by
a meromorphic function $\lambda$ and generates a trivial flow. $\lambda$
maps $Y$ to ${\Bbb {CP}}^1$. Hence the zeroes of the quasimomentum are
uniquely defined by the branch points and the branch points are the
only independent variables.

The Hamiltonian theory of the Whitham equations \cite{DN} is based on the
existence of the following Riemann metric on our moduli space:
\begin{displaymath}
ds^2=\sum_{i=1}^Ng_{ii}\left(d\lambda_{i} \right)^2, \hbox{ with }
\end{displaymath}
\begin{equation}
\label{metric}
g_{ii}= \hbox{res}\left.\vphantom{\frac{(dp)^2}{d\lambda}}\right|_{
\lambda=\lambda_i} \frac{(dp)^2}{d\lambda} \hbox{ for all } i=1,\ldots N, \
g_{ij}=0 \hbox{ for } i \ne j.
\end{equation}
(This formula is due to Dubrovin \cite{Du}).

In this section we show that isoperiodic flows are gradient flows in the
metric (\ref{metric}). This fact was pointed out to the authors by
S.P.Novikov. We shall consider the flows (\ref{FFMform}) where the
differential $dp$ is normalized by:
\begin{equation}
\oint_{a_i} dp=0, \ i=1,\ldots,g.
\end{equation}
Such flows preserve the $b$-periods of $dp$.

\begin{Theorem}
\label{grad-form}
Let $\omega_j$ be a holomorphic differential from the standard basis
(\ref{basis}).
Then the corresponding flow reads as:
\begin{equation}
\label{grad-eq}
\der{\lambda_k}{t_j}=g^{kl} \der {H_j}{\lambda_l},
\end{equation}
where
\begin{equation}
H_j=-\frac{1}{2\pi} \oint_{b_j} dp.
\end{equation}
(The metric tensor $g_{kl}$ and the inverse matrix $g^{kl}$ are diagonal
$g_{kl}=\delta^l_k g_{kk}$, the
corresponding Riemann metric is flat \cite{DN}.)

The isoperiodic property means that the functions $H_j$ are conserved
quantities for all flows
\begin{displaymath}
\der{H_j}{t_k}=0 \hbox{ for all } j,k.
\end{displaymath}
\end{Theorem}

The representation (\ref{grad-eq}) is very similar to the standard Hamiltonian
representation, but we have a symmetric form $g^{kl}$ instead of a
skew-symmetric one in the Hamilton theory.

In this example it is very important that this metric is not positive
definite. For a positive definite metric the function generating the flow
can not be a conservation law. But for the flows (\ref{grad-eq}) the
differentials of the functions $dH_j$, $j=1,\ldots,g$ lie in the `light
cone' of this metric. Thus all of them are conserved quantities for the
whole system. From this point of view equation (\ref{grad-eq}) is rather
similar to integrable Hamiltonian systems.

\begin{Remark}
Equation (\ref{grad-eq}) can be rewritten in a Hamiltonian form with the
same `Hamiltonians' but the corresponding skew-symmetric form has no
natural representation like (\ref{metric}).
\end{Remark}

\begin{Remark}
Introducing the flat coordinates of (\ref{metric}) we explicitly linearize
the system (\ref{grad-eq}). But the metric $g_{kk}$ has singularities and
the flat coordinates have rather complicated singularities in these points.
The metric $g_{kk}$ is nonsingular if and only if $dp$ and $d\lambda$ have
no common zeroes.
\end{Remark}

{\bf Proof of the Theorem \ref{grad-form}.}
Following the paper \cite{Du} we will
introduce flat coordinates for the metric (\ref{metric}).

For this purpose denote the order of the pole of the function $\lambda$ in the
point $y_k$ by $l_k+1$. It is natural to choose the local parameters $w_k$ in
the points $y_k$ so that
\begin{displaymath}
\lambda=w_k^{-l_k-1}.
\end{displaymath}

Consider the following collection of functions on our moduli space:
\begin{displaymath}
H_k=-\frac{1}{2\pi} \oint_{b_k} dp, \
t_k=\frac1{\sqrt{-1}} \oint_{a_k} p d \lambda, \ k=1,\ldots,g.
\end{displaymath}
\begin{displaymath}
r_k=\lim_{\gamma_1 \rightarrow y_1\atop \gamma_2 \rightarrow y_2}
\left [ \int_{\gamma_1}^{\gamma_2} dp + \frac1{\sqrt{-1}}f_1(\gamma_1)
-\frac1{\sqrt{-1}}f_k(\gamma_2) \right ],
\end{displaymath}
\begin{displaymath}
s_k=-\hbox{res}_{y_k} p d\lambda, \ k=2,\ldots,n.
\end{displaymath}
\begin{displaymath}
t_{\alpha,k}=-\hbox{res} \left. \vphantom{\frac{dp}{k(w_\alpha)^k}}
\right| _{y_\alpha} \frac{dp}{k(w_\alpha)^k}\ \alpha=1,\ldots,n,\
k=1,\ldots,l_\alpha.
\end{displaymath}

In the points of general position these functions give us a local coordinate
system. The metric (\ref{metric}) in these coordinates takes the form
\begin{eqnarray}
ds^2=\sum_{k=1}^g \left (dH_k dt_k + dt_k dH_k \right) +
\sum_{k=2}^n \left(dr_k ds_k+ds_k dr_k \right)+
\nonumber \\
+\sum_{\alpha=1}^n \left (l_\alpha+1 \right) \sum_{k=1}^{l_\alpha}
dt_{\alpha,k} dt_{\alpha,l_\alpha+1-k}.
\label{metric-const}
\end{eqnarray}

{}From (\ref{FFMform}) it follows that the flows generated
by holomorphic differentials do not change the variables $r_k$, $s_k$,
$t_{\alpha,k}$.

Comparison of (\ref{metric-const}) and (\ref{grad-eq})
finishes the proof.

\section{The moduli space of the
periodic NLS equation}
In this section we want to give a
description of the set of data
corresponding to periodic solutions of
the non-linear Schr{\"o}dinger equation.
It is quite easy to do the same for the
KdV equation.
As mentioned above the Riemann surfaces
corresponding to such periodic solutions
of the non-linear Schr{\"o}dinger
equation are described by the equation \begin{displaymath}
\mu ^{2}-\mu \Delta (\lambda )+1=0\end{displaymath}
with some entire function \begin{math}
\Delta \end{math}. This function may be
described as an infinite-sheeted covering
\begin{math} \Delta :{\Bbb {CP}}^{1}\rightarrow
{\Bbb {CP}}^{1}\end{math} with an essential
singularity at infinity. If the Riemann
surface is of finite genus\footnote{With
the help of the analysis of \cite{Sch}
our discussion may be extended to the
infinite genus case.} the function \begin{math}
\ln \mu \end{math} is meromorphic on both
sheets over some open
neighbourhood of \begin{math} \lambda =\infty
\end{math}. Hence the function
\begin{math} \arccos(\Delta /2)\end{math}
describes all sheets of the covering map
\begin{math} \Delta \end{math} over some
neighbourhood of \begin{math} \Delta =\infty
\end{math}. Furthermore, this local
coordinate near this branchpoint of
infinite order extends to all but a
finite number of the sheets over the
complete domain
\begin{math} \Delta \in {\Bbb {CP}}^{1}\end{math}.
Finally for big \begin{math} |\lambda |\end{math}
\begin{math} \, \ln \mu \end{math} is of the
form \begin{math} \pm \sqrt{-1}\lambda +{\bf
O}(1/\lambda )\end{math}. Hence the global
parameter \begin{math} \lambda \end{math} of
the covering space which covering is
induced by \begin{math} \Delta \end{math}
determined up to sign and summation of multiples of
\begin{math} 2\pi \sqrt{-1} \end{math} by this covering
map \begin{math} \Delta :{\Bbb {CP}}^{1}\rightarrow
{\Bbb {CP}}^{1}\end{math}. Later we will label the sheets of
this covering by the integers  and
will give a glueing rule for these sheets. The glueing rule
will uniquely determine the parameter \begin{math}
\lambda \end{math}. This observation proves the following
lemma:
\begin{Lemma} \label{Lemma3}
Let \begin{math} \lambda \mapsto \Delta (\lambda )\end{math}
correspond to a Riemann surface of
finite genus of a periodic solution of
the non-linear Schr{\"o}dinger equation.
Then this function is completely
determined by all the values of \begin{math}
\Delta \end{math} at the
branchpoints of the covering map \begin{displaymath}
{\Bbb C}\rightarrow {\Bbb C},\lambda \mapsto
\Delta (\lambda )\end{displaymath} together with
some glueing rule for the infinite number
of sheets.
\end{Lemma}
\hspace*{\fill } \begin{math} \Box \end{math}

\noindent
In the first section we introduced flow
parameters \begin{math} (t_{i})_{i\in {\Bbb
Z}}\end{math}, such that the values
of \begin{math} \Delta \end{math} at all
branchpoints of the covering map \begin{displaymath}
{\Bbb {CP}}^{1}\rightarrow {\Bbb {CP}}^{1},\lambda \mapsto
\Delta (\lambda )\end{displaymath} are equal to \begin{math}
(-1)^{i}2\cosh (t_{i})\end{math}. In
order to apply the previous Lemma
we have to add to these values the
glueing rule for the covering map. Let us
first consider the starting point, when
\begin{math} \Delta (\lambda )\end{math}  is equal
to \begin{math} 2\cos (\lambda )\end{math}.
In this case we choose the following
glueing rules: We have infinitely many
sheets labeled by some index \begin{math}
i\in {\Bbb Z}\end{math}. For all \begin{math}
j\in {\Bbb Z}\end{math} the \begin{math}
2j-1\end{math}-th sheet and the \begin{math}
2j\end{math}-th sheet has branchpoints
at \begin{math} \Delta =-2,\infty \end{math}
and the \begin{math} 2j\end{math}-th
sheet and the \begin{math}
2j+1\end{math}-th sheet has branchpoints
at \begin{math} \Delta =2,\infty \end{math},
and the cuts are chosen as indicated in
figure 1. The parameter \begin{math} \lambda \end{math}
is uniquely defined by the following choice:
On the \begin{math} 2j\end{math}-th sheet it is equal to
\begin{math} (2j-1)\pi \sqrt{-1} \end{math} for
\begin{math} \Delta =-2\end{math} and equal to
\begin{math} 2j\pi \sqrt{-1} \end{math} for
\begin{math} \Delta =2\end{math}.\\
On the \begin{math} 2j+1\end{math}-th sheet it is equal to
\begin{math} (2j+1)\pi \sqrt{-1} \end{math} for
\begin{math} \Delta =-2\end{math} and equal to
\begin{math} 2j\pi \sqrt{-1} \end{math} for
\begin{math} \Delta =2\end{math}.\\
\begin{picture}(396,150)(-190,-70)
\put(-140,-20){\circle*{2}}
\put(140,-20){\circle*{2}}
\put(-160,-30){\begin{math}
\Delta =-2\end{math} }
\put(120,-30){\begin{math}
\Delta =2\end{math} }
\put(-140,-20){\line(0,1){90} }
\put(140,-20){\line(0,1){90} }

\put(-185,10){
\begin{picture}(70,10)(0,0)
\put(0,35){\footnotesize cut between the}
\put(0,20){\footnotesize \begin{math} (2j-1)\end{math}-th and the}
\put(0,5){\footnotesize \begin{math} 2j\end{math}-th sheet}
\end{picture} }

\put(115,10){
\begin{picture}(70,10)(0,0)
\put(0,35){\footnotesize cut between the}
\put(0,20){\footnotesize \begin{math} 2j\end{math}-th and the}
\put(0,5){\footnotesize \begin{math} (2j+1)\end{math}-th sheet}
\end{picture} }

\put(-50,-65){Figure 1.}

\end{picture}

\noindent
With this glueing rule it is quite
obvious what happens, if two cuts along
a common sheet are passing each other.
In the following figures we move only
both cuts along the \begin{math}
2k\end{math}-th sheet for some fixed \begin{math}
k\in {\Bbb Z}\end{math} and assume that \begin{math}
j\end{math} runs through \begin{math}
{\Bbb Z}\setminus \{ k\}\end{math}. The
direction of the movement of the cuts from
one figure to the next figure is indicated by
arrows:\\
\begin{picture}(396,180)(-190,-80)

\put(-140,-20){\circle*{2}}
\put(140,-20){\circle*{2}}
\put(-160,-30){\begin{math}
\Delta =-2\end{math} }
\put(120,-30){\begin{math}
\Delta =2\end{math} }
\put(-140,-20){\line(0,1){95} }
\put(140,-20){\line(0,1){95} }

\put(-185,20){
\begin{picture}(70,50)(0,0)
\put(0,35){\footnotesize cut between the}
\put(0,20){\footnotesize \begin{math} (2j-1)\end{math}-th and the}
\put(0,5){\footnotesize \begin{math} 2j\end{math}-th sheet}
\end{picture} }

\put(115,20){
\begin{picture}(70,50)(0,0)
\put(0,35){\footnotesize cut between the}
\put(0,20){\footnotesize \begin{math} 2j\end{math}-th and the}
\put(0,5){\footnotesize \begin{math} (2j+1)\end{math}-th sheet}
\end{picture} }

\put(-70,-20){\circle*{2}}
\put(70,-20){\circle*{2}}
\put(-90,-30){\begin{math}
\Delta =-1\end{math} }
\put(50,-30){\begin{math}
\Delta =1\end{math} }
\put(-70,-20){\line(0,1){95} }
\put(70,-20){\line(0,1){95} }

\put(-90,20){
\begin{picture}(70,50)(0,0)
\put(0,35){\footnotesize cut between the}
\put(0,20){\footnotesize \begin{math} (2k-1)\end{math}-th and the}
\put(0,5){\footnotesize \begin{math} 2k\end{math}-th sheet}
\end{picture} }

\put(20,20){
\begin{picture}(70,50)(0,0)
\put(0,35){\footnotesize cut between the}
\put(0,20){\footnotesize \begin{math} 2k\end{math}-th and the}
\put(0,5){\footnotesize \begin{math} (2k+1)\end{math}-th sheet}
\end{picture} }

\put(-70,-20){\vector(2,1){70} }
\put(0,15){\vector(2,-1){70} }
\put(70,-20){\vector(-2,-1){70} }
\put(0,-55){\vector(-2,1){70} }

\put(-50,-75){Figure 2.}

\end{picture}

\noindent
\begin{picture}(396,180)(-190,-80)

\put(-140,-20){\circle*{2}}
\put(140,-20){\circle*{2}}
\put(-160,-30){\begin{math}
\Delta =-2\end{math} }
\put(120,-30){\begin{math}
\Delta =2\end{math} }
\put(-140,-20){\line(0,1){95} }
\put(140,-20){\line(0,1){95} }

\put(-185,20){
\begin{picture}(70,50)(0,0)
\put(0,35){\footnotesize cut between the}
\put(0,20){\footnotesize \begin{math} (2j-1)\end{math}-th and the}
\put(0,5){\footnotesize \begin{math} 2j\end{math}-th sheet}
\end{picture} }

\put(115,20){
\begin{picture}(70,50)(0,0)
\put(0,35){\footnotesize cut between the}
\put(0,20){\footnotesize \begin{math} 2j\end{math}-th and the}
\put(0,5){\footnotesize \begin{math} (2j+1)\end{math}-th sheet}
\end{picture} }

\put(-70,-20){\circle*{2}}
\put(70,-20){\circle*{2}}
\put(-90,-30){\begin{math}
\Delta =-1\end{math} }
\put(50,-30){\begin{math}
\Delta =1\end{math} }
\put(-70,-20){\line(0,1){95} }
\put(70,-20){\line(0,1){95} }

\put(-90,20){
\begin{picture}(70,50)(0,0)
\put(0,35){\footnotesize cut between the}
\put(0,20){\footnotesize \begin{math} 2k\end{math}-th and the}
\put(0,5){\footnotesize \begin{math} (2k+1)\end{math}-th sheet}
\end{picture} }

\put(20,20){
\begin{picture}(70,50)(0,0)
\put(0,35){\footnotesize cut between the}
\put(0,20){\footnotesize \begin{math} (2k-1)\end{math}-th and the}
\put(0,5){\footnotesize \begin{math} (2k+1)\end{math}-th sheet}
\end{picture} }

\put(-70,-20){\vector(2,1){70} }
\put(0,15){\vector(2,-1){70} }
\put(70,-20){\vector(-2,-1){70} }
\put(0,-55){\vector(-2,1){70} }

\put(-50,-75){Figure 3.}

\end{picture}

\noindent
\begin{picture}(396,180)(-190,-80)

\put(-140,-20){\circle*{2}}
\put(140,-20){\circle*{2}}
\put(-160,-30){\begin{math}
\Delta =-2\end{math} }
\put(120,-30){\begin{math}
\Delta =2\end{math} }
\put(-140,-20){\line(0,1){95} }
\put(140,-20){\line(0,1){95} }

\put(-185,20){
\begin{picture}(70,50)(0,0)
\put(0,35){\footnotesize cut between the}
\put(0,20){\footnotesize \begin{math} (2j-1)\end{math}-th and the}
\put(0,5){\footnotesize \begin{math} 2j\end{math}-th sheet}
\end{picture} }

\put(115,20){
\begin{picture}(70,50)(0,0)
\put(0,35){\footnotesize cut between the}
\put(0,20){\footnotesize \begin{math} 2j\end{math}-th and the}
\put(0,5){\footnotesize \begin{math} (2j+1)\end{math}-th sheet}
\end{picture} }
\put(-70,-20){\circle*{2}}
\put(70,-20){\circle*{2}}
\put(-90,-30){\begin{math}
\Delta =-1\end{math} }
\put(50,-30){\begin{math}
\Delta =1\end{math} }
\put(-70,-20){\line(0,1){95} }
\put(70,-20){\line(0,1){95} }

\put(-90,20){
\begin{picture}(70,50)(0,0)
\put(0,35){\footnotesize cut between the}
\put(0,20){\footnotesize \begin{math} (2k-1)\end{math}-th and the}
\put(0,5){\footnotesize \begin{math} (2k+1)\end{math}-th sheet}
\end{picture} }

\put(20,20){
\begin{picture}(70,50)(0,0)
\put(0,35){\footnotesize cut between the}
\put(0,20){\footnotesize \begin{math} (2k-1)\end{math}-th and the}
\put(0,5){\footnotesize \begin{math} 2k\end{math}-th sheet}
\end{picture} }

\put(-50,-75){Figure 4.}

\end{picture}

\noindent
{}From the figures 2. to the figure 4. the
both cuts along the \begin{math}
2k\end{math}-th sheet have moved once
around each other. Finally the
indexes \begin{math} (2k-1,2k,2k+1)\end{math}
 are permuted to \begin{math}
(2k+1,2k-1,2k)\end{math}. This shows
that the coordinate given by the values
of \begin{math} \Delta \end{math}  at the
branchpoints has a singularity of the
form \begin{math}
(\Delta _{2k-1}-\Delta _{2k})^{1/3}\end{math},
where \begin{math} \Delta _{2k}=(-1)^{2k-1}2\cosh
(t_{2k})\end{math}
denotes at the starting point the
value of \begin{math} \Delta \end{math}
at the branchpoint between the \begin{math}
2k-1\end{math}-th and the \begin{math}
2k\end{math}-th sheet and \begin{math}
\Delta _{2k}=(-1)^{2k}2\cosh
(t_{2k})\end{math} denotes at the
starting point the
value of \begin{math} \Delta \end{math}
at the branchpoint between
the \begin{math} 2k\end{math}-th and the
\begin{math} 2k+1\end{math}-th sheet.
This gives a complete picture of the
branchpoints of the coordinates \begin{math}
(t_{i})_{i\in {\Bbb Z}}\end{math}.

Let us now describe the structure of the covering
of the moduli space with respect to
these coordinates. For reasons of
simplicity let us assume that the real
parts of the values of \begin{math} \Delta \end{math}
at two branchpoints are different, whenever
they belong to a common sheet. In this
case all the cuts may be chosen of the
form \begin{math} \Delta _{i}+\sqrt{-1} [0,\infty
]\end{math}. Hence we have to associate
to the values of \begin{math} \Delta \end{math}
at all the branchpoints the numbers of
the two corresponding sheets. Due to the
previous Lemma this completely determines
the corresponding data. These data may
be described by a graph with vertices
indexed by \begin{math} i\in {\Bbb Z}\end{math}
corresponding to the sheets and links
corresponding to the branchpoints and a
function from the links into the complex
numbers, which is equal to the value of \begin{math}
\Delta \end{math} at the
corresponding branchpoint. These data
have to obey the following conditions:
\begin{description}
\item [(i)] The graph is connected.
\item[(ii)] All but finitely many
vertices are linked exactly with those
vertices, whose index differs by \begin{math}
\pm 1\end{math}.
\item[(iii)] The number of links is
equal to the number of pairs of the form
\begin{math} (i,i+1)\mbox{ with }i\in {\Bbb Z}
\end{math}. Due to
condition (ii) this has a precise
meaning.
\item[(iv)] For large \begin{math} |i|\end{math}
the number corresponding to the link
connecting the \begin{math}
i\end{math}-th and the \begin{math}
i+1\end{math}-th vertex is equal to\footnote{In the
infinite genus case this condition has
to be weakened to the condition that the
number is almost equal to \begin{math}
(-1)^{i}2\end{math} in a sense, which
depends on the class of solutions, which
are considered (see \cite{Sch}).}
\begin{math}
(-1)^{i}2\end{math}
\end{description}
It is quite obvious how this description
may be extended to the general case.
In the following figures we draw the
graphs corresponding to figures 1-4 with
k=0:\\
\begin{picture}(396,100)(-210,-50)

\put(-150,0){\begin{math} \ldots \end{math} }
\put(-100,0){\circle*{2}}
\put(-50,0){\circle*{2}}
\put(0,0){\circle*{2}}
\put(50,0){\circle*{2}}
\put(100,0){\circle*{2}}
\put(150,0){\begin{math} \ldots \end{math} }

\put(-100,0){\line(-1,0){25} }
\put(-50,0){\line(-1,0){50} }
\put(0,0){\line(-1,0){50} }
\put(50,0){\line(-1,0){50} }
\put(100,0){\line(-1,0){50} }
\put(100,0){\line(1,0){25} }

\put(-105,-10){-2}
\put(-55,-10){-1}
\put(0,-10){0}
\put(45,-10){1}
\put(95,-10){2}

\put(-130,2){-2}
\put(-80,2){2}
\put(-30,2){-2}
\put(20,2){2}
\put(70,2){-2}
\put(120,2){2}

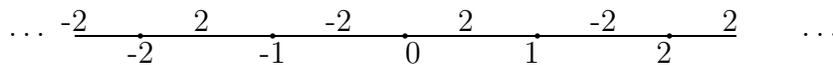
\put(-100,-45){Graph corresponding to Figure 1.}

\end{picture}

\noindent
\begin{picture}(396,100)(-210,-50)

\put(-150,0){\begin{math} \ldots \end{math} }
\put(-100,0){\circle*{2}}
\put(-50,0){\circle*{2}}
\put(0,0){\circle*{2}}
\put(50,0){\circle*{2}}
\put(100,0){\circle*{2}}
\put(150,0){\begin{math} \ldots \end{math} }

\put(-100,0){\line(-1,0){25} }
\put(-50,0){\line(-1,0){50} }
\put(0,0){\line(-1,0){50} }
\put(50,0){\line(-1,0){50} }
\put(100,0){\line(-1,0){50} }
\put(100,0){\line(1,0){25} }

\put(-105,-10){-2}
\put(-55,-10){-1}
\put(0,-10){0}
\put(45,-10){1}
\put(95,-10){2}

\put(-130,2){-2}
\put(-80,2){2}
\put(-30,2){-1}
\put(20,2){1}
\put(70,2){-2}
\put(120,2){2}

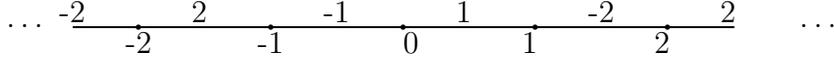
\put(-100,-45){Graph corresponding to Figure 2.}

\end{picture}

\noindent
\begin{picture}(396,100)(-210,-50)

\put(-150,0){\begin{math} \ldots \end{math} }
\put(-100,0){\circle*{2}}
\put(-50,10){\circle*{2}}
\put(0,-10){\circle*{2}}
\put(50,10){\circle*{2}}
\put(100,0){\circle*{2}}
\put(150,0){\begin{math} \ldots \end{math} }

\put(-100,0){\line(-1,0){25} }
\put(-50,10){\line(-5,-1){50} }
\put(0,-10){\line(5,2){50} }
\put(50,10){\line(-1,0){100} }
\put(100,0){\line(-5,1){50} }
\put(100,0){\line(1,0){25} }

\put(-105,-10){-2}
\put(-55,0){-1}
\put(0,-20){0}
\put(45,0){1}
\put(95,-10){2}

\put(-130,2){-2}
\put(-80,7){2}
\put(-5,12){1}
\put(20,-10){-1}
\put(70,7){-2}
\put(120,2){2}

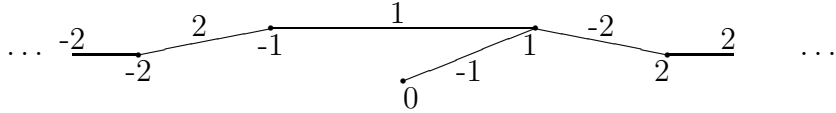
\put(-100,-45){Graph corresponding to Figure 3.}

\end{picture}

\noindent
\begin{picture}(396,100)(-210,-50)

\put(-150,0){\begin{math} \ldots \end{math} }
\put(-100,0){\circle*{2}}
\put(-50,10){\circle*{2}}
\put(0,-10){\circle*{2}}
\put(50,10){\circle*{2}}
\put(100,0){\circle*{2}}
\put(150,0){\begin{math} \ldots \end{math} }

\put(-100,0){\line(-1,0){25} }
\put(-50,10){\line(-5,-1){50} }
\put(0,-10){\line(-5,2){50} }
\put(50,10){\line(-1,0){100} }
\put(100,0){\line(-5,1){50} }
\put(100,0){\line(1,0){25} }

\put(-105,-10){-2}
\put(-55,0){-1}
\put(0,-20){0}
\put(45,0){1}
\put(95,-10){2}

\put(-130,2){-2}
\put(-80,7){2}
\put(-5,12){-1}
\put(-30,-10){1}
\put(70,7){-2}
\put(120,2){2}

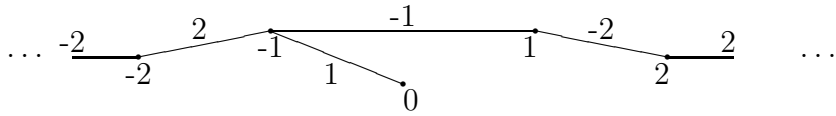
\put(-100,-45){Graph corresponding to Figure 4.}

\end{picture}

\noindent
Any neighbourhood of the point \begin{math} \lambda =\infty \end{math}
may be described by glueing together parts of finitely many sheets over
some neighbourhoods of the point \begin{math}
\Delta =\infty \end{math} with all other complete sheets over \begin{math}
\Delta \in {\Bbb {CP}}^{1}\end{math}. Now \begin{math} \sqrt{-1}\ln \mu =
\arccos(\Delta /2)\end{math}
defines a single-valued function on some of these neighbourhoods.
Hence the glueing rules of all the parts of the sheets of some
of these neighbourhoods must coincide with the corresponding
glueing rules described by the standard graph. Now let us
consider paths in the \begin{math} \lambda \end{math}-plane, which
covers big circles in the \begin{math} \Delta \end{math}-plane
moving clockwise around. They fit together to two paths with
open ends. The corresponding succesion of those sheets, in which
the path crosses the negative imaginary axis of the
\begin{math} \Delta \end{math}-plane (and therefore the main
part of the corresponding turn),
results in two sequences of labels of our sheets:
\begin{displaymath} \ldots ,2n+2,2n,2n-2,\ldots \mbox{ and }
\ldots , 2n-1,2n+1,2n+3,\ldots \end{displaymath}
Hence for all graphs, which corresponds to periodic solutions
of the non-linear Schr{\"o}dinger equation, these paths
should result in these two sequences of labels of our sheets.
It is obvious that each part of these paths, which corresponds to
one big circle moving clockwise around from the negative imaginary
axis back to the negative imaginary axis in the
\begin{math} \Delta \end{math}-plane results in a subgraph
of the form indicated in figure 5, such that the following conditions
are fulfilled:
\begin{description}
\item[(a)] The real parts of the values of \begin{math}
\Delta \end{math} at the branchpoints are ordered
\begin{math} \hbox{Re}(\Delta_{1})<\ldots <\hbox{Re}(\Delta_{l})\end{math}.
\item[(b)] Any sheet with label \begin{math}k_{j},j=1,\ldots ,
l-1\end{math} has no branchpoint, such that the real part of the
corresponding value of \begin{math} \Delta \end{math} lies
between \begin{math}
\hbox{Re}(\Delta_{j})\mbox{ and }\hbox{Re}(\Delta_{j+1})\end{math}.
\item[(c)] The subgraph is maximal with respect to conditions
(a) and (b). This means that the sheet with label
\begin{math}k_{0}\end{math} has no branchpoint,
such that the real part of the corresponding value of
\begin{math} \Delta \end{math} is less than
\begin{math} \hbox{Re}(\Delta_{1})\end{math} and the sheet with label
\begin{math}k_{l}\end{math} has no branchpoint,
such that the real part of the corresponding value of
\begin{math} \Delta \end{math} is bigger than
\begin{math} \hbox{Re}(\Delta_{l})\end{math}.
\end{description}
\begin{picture}(396,100)(-210,-50)

\put(-100,0){\circle*{2}}
\put(-25,0){\circle*{2}}
\put(50,0){\circle*{2}}
\put(125,0){\circle*{2}}

\put(-25,0){\line(-1,0){75} }
\put(125,0){\line(-1,0){75} }
\put(-25,0){\line(1,0){20} }
\put(50,0){\line(-1,0){20} }
\put(0,0){\ldots }

\put(-105,-10){\begin{math} k_{0}\end{math} }
\put(-30,-10){\begin{math} k_{1}\end{math} }
\put(45,-10){\begin{math} k_{l-1}\end{math} }
\put(120,-10){\begin{math} k_{l}\end{math} }

\put(-75,2){\begin{math} \Delta_{1}\end{math} }
\put(75,2){\begin{math} \Delta_{l}\end{math} }

\put(-50,-45){Figure 5.}

\end{picture}

\noindent
This gives the following condition on our graphs:
\begin{description}
\item[(v)] For each subgraph, which obeys
condition (a), (b) and (c) the labels of the endpoints
obey the equation
\begin{math} k_{0}-k_{l}=2(-1)^{k_{0}}=2(-1)^{k_{l}}\end{math}.
\end{description}
Now we can state the main theorem of
this section:
{\tolerance=10000 \begin{Theorem} \label{theorem4}
The graphs described above are in one to
one correspondence with the data, which
corresponds to periodic solutions of the
non-linear Schr{\"o}dinger equation.
\end{Theorem}

}
Proof: Let us first prove that to all
such graphs there exists some
entire function \begin{math} {\Bbb C}\rightarrow
{\Bbb C},\lambda \mapsto \Delta (\lambda )\end{math}.
In order to do this we divide the
infinite number of cuts in three parts:
To any part of the graphs of the form
indicated in figure 6. and figure 7. there
corresponds the entire function \begin{displaymath}
{\Bbb C}\rightarrow {\Bbb C},
\lambda \mapsto (-1)^{k}2 \cos (\sqrt{\lambda })\end{displaymath}
\begin{picture}(396,100)(-180,-50)

\put(-100,0){\circle*{2}}
\put(-25,0){\circle*{2}}
\put(50,0){\circle*{2}}
\put(125,0){\circle*{2}}
\put(200,0){\begin{math} \ldots \end{math} }

\put(-25,0){\line(-1,0){75} }
\put(50,0){\line(-1,0){75} }
\put(125,0){\line(-1,0){75} }
\put(125,0){\line(1,0){40} }

\put(-105,-10){k}
\put(-30,-10){k+1}
\put(45,-10){k+2}
\put(120,-10){k+3}

\put(-85,2){\begin{math} (-1)^{k}2\end{math} }
\put(-10,2){\begin{math} (-1)^{k+1}2\end{math} }
\put(65,2){\begin{math} (-1)^{k}2\end{math} }
\put(140,2){\begin{math} (-1)^{k+1}2\end{math} }

\put(-50,-45){Figure 6.}

\end{picture}

\noindent
\begin{picture}(396,100)(-210,-50)

\put(-200,0){\begin{math} \ldots \end{math} }
\put(-125,0){\circle*{2}}
\put(-50,0){\circle*{2}}
\put(25,0){\circle*{2}}
\put(100,0){\circle*{2}}

\put(-125,0){\line(-1,0){40} }
\put(-50,0){\line(-1,0){75} }
\put(25,0){\line(-1,0){75} }
\put(100,0){\line(-1,0){75} }

\put(-130,-10){k-2}
\put(-55,-10){k-1}
\put(20,-10){k}
\put(95,-10){k+1}

\put(-180,2){\begin{math} (-1)^{k-1}s\end{math} }
\put(-105,2){\begin{math} (-1)^{k}2\end{math} }
\put(-35,2){\begin{math} (-1)^{k-1}2\end{math} }
\put(40,2){\begin{math} (-1)^{k}2\end{math} }

\put(-50,-45){Figure 7.}

\end{picture}

\noindent
Then due to conditions (ii) and (iv) the total
covering space corresponding to some
graph may be obtained by glueing
in the middle of two such
copies of \begin{math} {\Bbb {CP}}^{1}\end{math}
a finite number of copies of \begin{math}
{\Bbb {CP}}^{1}\end{math} along some cuts
described by the glueing rule given by
the graph. Due to conditions (i) and (iii) this
total covering space is isomorphic to \begin{math}
{\Bbb {CP}}^{1}\end{math}. This shows that all
these graphs define some infinite sheeted
covering \begin{math} {\Bbb {CP}}^{1}\rightarrow
{\Bbb {CP}}^{1}\end{math} induced by some
entire function \begin{math} \lambda \mapsto
\Delta (\lambda )\end{math}, where the parameter \begin{math}
\lambda \end{math} is defined only up to some
rational transformation, which leaves invariant
the point \begin{math} \infty \end{math}. The
transcendental curve defined by the
equation
\begin{math} \mu ^{2}-\mu \Delta (\lambda )+1=0\end{math}
has infinitely many ordinary double
points. In fact, all the branchpoints
of order 1 of the covering map
defined by these graphs, at which \begin{math}
\Delta \end{math} is equal to \begin{math}
\pm 2\end{math} correspond to such an
ordinary double point. Hence the
normalization of this transcendental
curve is a compact Riemann surface. Due to condition
(v) the function \begin{math} 1/\arccos (\Delta /2)\end{math}
is a local parameter in the covering space near
infinity. Hence there exists a unique global
parameter \begin{math} \lambda \end{math} of the
covering space, such that \begin{math}
\lambda = \arccos (\Delta /2) + {\bf O}(1/\lambda )\end{math}
near infinity. This shows that the Riemann surface
defined by the equation
\begin{math} \mu ^{2}-\mu \Delta (\lambda )+1=0\end{math}
corresponds to periodic solutions of the non-linear
Schr{\"o}dinger equation. The other direction of the one
to one correspondence is part of the
definition of the graph.
\hspace*{\fill } \begin{math} \Box \end{math}
\begin{Corollary}
The domain of the flow parameters
\begin{math} (t_{i})_{i\in {\Bbb
Z}}\end{math} of the non-linear
Schr{\"o}dinger equation, defined in the
first section may be chosen to be equal
to \begin{displaymath} \left\{ \left. (t_{i})_{i\in {\Bbb
Z}}\in {\Bbb C}^{{\Bbb Z}} \right| \mbox{
all but a finite number of \begin{math}
t_{i}\end{math}'s are eqal to zero} \right\}
.\end{displaymath} More precisely, on this domain these
flows have have only branchpoints of
finite oder.
\end{Corollary}
\hspace*{\fill } \begin{math} \Box \end{math}

\noindent
In the proof we described implicitly how to
determine the genus of the Riemann surface from the
graph. In fact, to all branchpoints of the
covering map induced by \begin{math} \Delta
\end{math} there correspond two branchpoints
of the covering map induced by \begin{math} \mu
\end{math} from \begin{math} Y \rightarrow
{\Bbb {CP}}^{1}\end{math}. But those branchpoints,
where \begin{math} \Delta \end{math} is equal to
\begin{math} \pm 2\end{math} correspond to ordinary
double points and may be neglected, if they have
no common sheet.

The non-linear Schr{\"o}dinger equation
has two natural reality conditions.
For both of them the function \begin{math} \Delta
(\lambda )\end{math} is assumed to be real, for all
\begin{math} \lambda \in {\Bbb R}\end{math}. This
condition implies that \begin{math} \overline{\Delta
(\bar{\lambda } )}\end{math} is equal to \begin{math}
\Delta (\lambda )\end{math}. This gives the existence of
an antilinear involution of the covering space of the
covering map induced by \begin{math} \lambda \rightarrow
\Delta (\lambda )\end{math} corresponding to the antilinear
involution \begin{math} \Delta \rightarrow
\bar{\Delta }\end{math}. Hence the cuts must be chosen to be
invariant under this involution. For the starting
point this may be achieved by turning the cuts:\\
\begin{picture}(396,100)(-200,-20)

\put(-80,20){\circle*{2}}
\put(80,20){\circle*{2}}
\put(-100,10){\begin{math}
\Delta =-2\end{math} }
\put(60,10){\begin{math}
\Delta =2\end{math} }
\put(-80,20){\line(-1,0){110} }
\put(80,20){\line(1,0){110} }

\put(-180,20){
\begin{picture}(70,50)(0,0)
\put(0,35){\footnotesize cut between the}
\put(0,20){\footnotesize \begin{math} (2j-1)\end{math}-th and the}
\put(0,5){\footnotesize \begin{math} 2j\end{math}-th sheet}
\end{picture} }

\put(100,20){
\begin{picture}(70,50)(0,0)
\put(0,35){\footnotesize cut between the}
\put(0,20){\footnotesize \begin{math} 2j\end{math}-th and the}
\put(0,5){\footnotesize \begin{math} (2j+1)\end{math}-th sheet}
\end{picture} }

\put(-50,-15){Figure 1'.}

\end{picture}

\noindent
There are more complicated situations, in which this
antilinear involution permutes some of the sheets and
the values of \begin{math} \Delta \end{math} are not real.
In this case all but a finite number of cuts may be
chosen to be invariant under the antilinear involution
and therefore are part of the real axis \begin{math}
\Delta \in {\Bbb R}\end{math}. But finitely many cuts
are interchanged under this antilinear involution. We
may choose them to be of the form \begin{displaymath}
\Delta =\Delta _{0}+\sqrt{-1} {\Bbb R}_{+}\mbox{ and }
\Delta =\bar{\Delta _{0}}-\sqrt{-1} {\Bbb R}_{+}
\mbox{ with } \hbox{Im}{\Delta _{0}} \geq 0\mbox{, respectively.}
\end{displaymath}
The antilinear involution is then described by an involution
of the sheets and the branchpoints, such that the
value of \begin{math} \Delta \end{math} at
two branchpoints, which are interchanged under
this involution are the complex conjugate of each other.
Let us give an example. In the following figure
we describe the cuts by the pairs of the corresponding
numbers of sheets. Again some number \begin{math} k
\in {\Bbb Z}\end{math} is fixed:\\
\begin{picture}(396,195)(-200,-105)

\put(-95,0){\circle*{2}}
\put(95,0){\circle*{2}}
\put(-120,-15){\begin{math}
\Delta =-2\end{math} }
\put(80,-15){\begin{math}
\Delta =2\end{math} }
\put(-95,0){\line(-1,0){100} }
\put(95,0){\line(1,0){100} }

\put(-180,8){\footnotesize
\begin{math} (2j-1,2j)\mbox{ with } j\neq k\end{math}
}
\put(60,18){\footnotesize
\begin{math} (2j,2j+1)\mbox{ with } j\neq k,k+1\end{math}
}
\put(90,8){\footnotesize
and \begin{math} (2k-2,2k+1)\end{math}
}

\put(0,25){\circle*{2}}
\put(0,-25){\circle*{2}}
\put(-30,15){\begin{math}
\Delta =.5\sqrt{-1} \end{math} }
\put(-30,-22){\begin{math}
\Delta =-.5\sqrt{-1} \end{math} }
\put(0,25){\line(0,1){60} }
\put(0,-25){\line(0,-1){60} }

\put(-25,45){\footnotesize
\begin{math} (2k,2k+1)\end{math}
}
\put(-30,-55){\footnotesize
\begin{math} (2k-1,2k+1)\end{math}
}

\put(-50,-100){Figure 8.}

\end{picture}

\noindent
The involution of the sheets is given by the map
\begin{displaymath} {\Bbb Z} \rightarrow {\Bbb Z},
i\mapsto i\mbox{ for } i\neq k-1,k;\ k-1\mapsto k;
\mbox{ and }k\mapsto k-1.
\end{displaymath}
Again we may combine this glueing rule to a graph:\\
\begin{picture}(396,130)(-210,-70)

\put(-140,0){\begin{math} \ldots \end{math} }
\put(-70,0){\circle*{2}}
\put(0,0){\circle*{2}}
\put(70,0){\circle*{2}}
\put(0,40){\circle*{2}}
\put(0,-40){\circle*{2}}
\put(140,0){\begin{math} \ldots \end{math} }

\put(-70,0){\line(-1,0){40} }
\put(0,0){\line(-1,0){70} }
\put(70,0){\line(-1,0){70} }
\put(70,0){\line(1,0){40} }
\put(0,0){\line(0,1){40} }
\put(0,0){\line(0,-1){40} }

\put(-80,-10){2k-2}
\put(-10,-10){2k+1}
\put(60,-10){2k+2}
\put(-7,42){2k}
\put(-10,-50){2k-1}

\put(-115,2){-2}
\put(-45,2){2}
\put(25,2){-2}
\put(95,2){2}
\put(-40,20){\begin{math} .5\sqrt{-1} \end{math} }
\put(-45,-30){\begin{math} -.5\sqrt{-1} \end{math} }

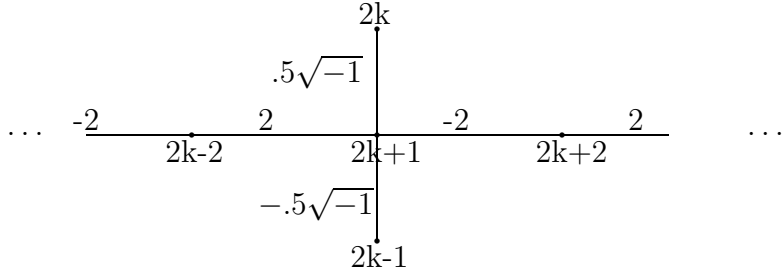
\put(-100,-65){Graph corresponding to Figure 8.}

\end{picture}

\noindent
Now we can describe the restrictions
of the two reality conditions:
The so called focusing case corresponds
to the restriction that the values of \begin{math}
\Delta \end{math} at the branchpoints are
elements of \begin{math} [-\infty
,-2]\cup [2,\infty ]\end{math}. In this case
the links and vertices of the graph is always
of the same form as the graph corresponding to
figure 1. Then it is obvious that
the coordinates \begin{math} (t_{i})_{i\in
{\Bbb Z}} \end{math} are global single
valued coordinates in the focusing case.
The so called defocusing case corresponds
to the restriction that the function
\begin{math} \Delta (\lambda )\end{math} takes values
in \begin{math} [-2,2]\end{math} for all
\begin{math} \lambda \in {\Bbb R} \end{math}.
Hence the corresponding graph have to consist
of one connected purely horizontal part
combining all sheets and branchpoints, which are
invariant under the involution describing the antilinear
involution of the covering space of \begin{math} \Delta
\end{math}. The corresponding values \begin{math} \Delta
\end{math} have to be elements of
\begin{math} [-2,2]\end{math} and have to be
alternating with respect to the order.
Beside this horizontal part the graph may have
finitely many vertical strings of finite length, which
are symmetric with respect to the horizontal
part. This symmetry implements the antilinear
involution. Hence the values of \begin{math}
\Delta \end{math} above the horizontal part of
the graph are assumed to have non-negative
imaginary part and to be the complex conjugate
of the corresponding values below the horizontal
part. In this case
the coordinates \begin{math} (t_{i})_{i\in
{\Bbb Z}} \end{math}
are multivalued with complicated
branchpoints.

We want to remark that \begin{math}
d\lambda \end{math} has a zero at the points
where \begin{math} \Delta ^{2}-4\end{math} is
equal to zero and \begin{math} d\mu \end{math}
has a zeros at the branchpoints of the
covering map induced by \begin{math}
\Delta \end{math}. But all the ordinary
double points may be neglected. Hence on
the normalization of the algebraic curve
defined by \begin{math} \mu ^{2}-\mu \Delta +1=0\end{math}
the differentials \begin{math} d\lambda \end{math}
and \begin{math} d\mu \end{math} have a
common zero, whenever the values of \begin{math}
\Delta \end{math} at two branchpoints
of a common sheet are equal to \begin{math}
\pm2\end{math}.

Finally let us mention that
the differential equations describing the flows
to the flow parameters \begin{math} (t_{i})_{i\in {\Bbb Z}}
\end{math} can be solved numerically. We implemented these algebraic
differential equations with the help of
standard programs up to genus equal to 6.
In particular, we checked the branchpoints of
these flow parameters numerically. With the help
of such a program one can calculate quite quickly all the
Riemann surfaces corresponding to the graphs described
above.

\end{document}